\documentclass{emulateapj}

\usepackage{amssymb}
\usepackage{apjfonts}

\usepackage{lscape}
\usepackage{epsfig}
\usepackage{times}
\usepackage{natbib}
\bibpunct{(}{)}{,}{a}{}{}

\shorttitle{Deep SWIRE Field IV. The sub--mJy galaxy population}
\shortauthors{Strazzullo et al.}

\begin{document}


\title{The Deep Swire Field. IV. First properties of the sub--\lowercase{m}J\lowercase{y}
galaxy population:\\ redshift distribution, AGN activity and star formation.}


\author{Veronica Strazzullo\altaffilmark{1}, Maurilio Pannella\altaffilmark{1},
Frazer N. Owen\altaffilmark{1},\\ Ralf Bender\altaffilmark{2,3}, Glenn
E. Morrison\altaffilmark{4,5}, Wei-Hao Wang\altaffilmark{6,7}, David L. Shupe\altaffilmark{8}   }

\altaffiltext{1}{National Radio Astronomy Observatory, 1003 Lopezville Rd.,
  Socorro, NM 87801, USA }
\altaffiltext{2}{Universit{\"a}ts-Sternwarte, Scheinerstrasse 1, Munich D-81679,
  Germany}
\altaffiltext{3}{Max-Planck-Institut f{\"u}r Extraterrestrische Physik, Giessenbachstrasse, 85748 Garching-bei-M{\"u}nchen, Germany}
\altaffiltext{4}{Institute for Astronomy, University of Hawaii, Honolulu,
  Hawaii, 96822, USA}
\altaffiltext{5}{Canada-France-Hawaii Telescope, Kamuela, Hawaii, 96743, USA}
\altaffiltext{6}{Institute of Astronomy and Astrophysics, Academia
Sinica, P.O. Box 23-141,Taipei 10617, Taiwan}
\altaffiltext{7}{Leung Center for Cosmology and Particle Astrophysics,
  National Taiwan University, No. 1, Sec. 4, Roosevelt Road, Taipei 10617, Taiwan}
\altaffiltext{8}{Infrared Processing and Analysis Center, California Institute of Technology, 100-22, Pasadena, CA, 91125, USA}


\begin{abstract}

We present a study of a 20cm selected sample in the Deep SWIRE VLA
Field, reaching a 5--$\sigma$ limiting flux density at the image
center of $S_{1.4GHz}\sim13.5 \mu$Jy. In a $0.6\times0.6$ square
degrees field, we are able to assign an optical/IR counterpart to 97\%
of the radio sources.  Up to 11 passbands from the NUV to 4.5$\mu$m
are then used to sample the spectral energy distribution (SED) of
these counterparts in order to investigate the nature of the host
galaxies. By means of an SED template library and stellar population
synthesis models we estimate photometric redshifts, stellar masses,
and stellar population properties, dividing the sample in three
sub--classes of quiescent, intermediate and star--forming galaxies. We
focus on the radio sample in the redshift range $0.3<z<1.3$ where we
estimate to have a redshift completeness higher than 90\%, and study
the properties and redshift evolution of these sub--populations.  We
find that, as expected, the relative contributions of AGN and
star--forming galaxies to the $\mu$Jy population depend on the flux
density limit of the sample. At all flux levels a significant population of
``green--valley'' galaxies is observed.  While the actual nature of
these sources is not definitely understood, the results of this work
may suggest that a significant fraction of faint radio sources might be
composite (and possibly transition) objects, thus a simple ``AGN vs
star--forming'' classification might not be appropriate to fully
understand what faint radio populations really are.

\end{abstract}





\section{Introduction}
\label{intro}

In recent years, many studies have agreed in assigning a relevant role
to active galactic nuclei (AGN) feedback in shaping the evolution of
galaxies, and in particular their star formation histories, making the
co--evolution of galaxies and AGNs a fundamental piece in the puzzle
of the general evolution of galaxy populations \citep[e.g.][]{croton2006,menci2006,bower2006,monaco2007,somerville2008}. As it is now believed,
basically all massive galaxies in the local Universe harbor a massive
black hole, and the correlation between black--hole mass and galaxy
bulge mass \citep[e.g.][]{KormendyGebhardt2001} points toward a close
link between the formation of the black hole and of its host galaxy.

At the same time, deep radio surveys have been conducted in
association with multi--wavelength observations, allowing such
(co--)evolution of galaxies and massive black holes to be probed.
These deep radio surveys, for the most part at 1.4 GHz, opened a
window on the previously largely unexplored $\mu Jy$ populations.
However, unlike the Jy and mJy populations, which are dominated by
radio loud AGNs hosted by quiescent galaxies, the $\mu Jy$ radio
source population appears to be increasingly dominated by different
kind of sources, star--forming galaxies and low--luminosity AGNs \citep[e.g. among others][]{windhorst1985,condon1989,jarviserawlings2004}.

Beside being studied at radio wavelengths
\citep[e.g.][]{ciliegi1999,richards2000,bondi2003,hopkins2003,huynh2005},
the dual nature of this composite $\mu Jy$ populations has also been
confirmed with X--ray and far--infrared observations
\citep[e.g.][]{afonso2001, afonso2006,georgakakis2003,georgakakis2004}.
Nonetheless, the individual contribution of AGNs and star--forming
galaxies to the whole $\mu Jy$ population has proved difficult to 
determine accurately for several reasons, including the often small
size of the samples, as well as observational biases introduced, for
instance, by optical (and in particular, but not only, spectroscopic)
identification of the counterparts and follow--up.  Needless to say,
this is even more true at higher redshifts, thus hampering our ability
to set evolutionary constraints.  Therefore, while many studies over
several years have been devoted to this investigation, making use of
different kinds of information at different wavelengths \citep[e.g.][]{windhorst1985,georgakakis1999,gruppioni1999,richards1999,ciliegi2003,gruppioni2003,seymour2004,cowie2004,afonso2005,huynh2005,afonso2006,simpson2006,fomalont2006,barger2007,seymour2008,ibar2008,ibar2009,bardelli2009},
they sometimes have produced controversial results.

In spite of these difficulties deep radio surveys have been
recognized, for both the AGN and star--forming components, as an
exceptional, powerful tool, even though their potential is not yet
fully exploited. First, since radio emission is basically unaffected
by dust extinction, important issues at optical wavelengths,
e.g. obscured star formation and highly obscured AGNs missing from
deep X-ray surveys \citep[but see discussions in e.g.][and references
therein]{barger2007,tozzi2009}, clearly find a solution when observing
at radio wavelengths.  In fact, radio--selected AGN samples are not
the same as AGN samples selected at other wavelengths, since they
include populations of low--power radio sources which would not be
classified as AGNs from their optical or X--ray properties
\citep[e.g.][]{best2005b,hardcastle2006,hickox2009}, pointing toward an
intrinsically different nature of these sources. Furthermore, the
arcsecond resolution available for some of these radio surveys makes
it relatively easy to cross--correlate them with other data across a
broad wavelength range including optical and near--infrared. This is
actually a fundamental point, because in fact the study of the faint
radio sources at many different wavelengths obviously maximizes the
scientific return of the radio survey, allowing a more complete
characterization of a population which is intrinsically mixed at radio
wavelengths. In particular, the cross--correlation with large
X--ray/optical/NIR surveys where a wealth of information is available
in terms of spectroscopic/photometric redshifts, stellar populations
and galaxy morphologies, enhances our understanding of the nature of
these sources, out to redshift $\approx 1$. Just in the last couple of
years, several studies were published making use of such deep,
panchromatic observations in order to investigate the different galaxy
species populating the $\mu Jy$ samples, as for instance
\citet{smolcic2008} in the COSMOS field , \citet{mainieri2008} and
\citet{padovani2009} in the GOODS--CDFS field, and \citet{huynh2008}
in the HDF--S.

This paper is the fourth in a series documenting our study of the
deep SWIRE field centered at 10$^h$46$^m$00$^s$, 
59\arcdeg01\arcmin00\arcsec (J2000). Paper I \citep{SDFI} describes
the 20cm VLA observations which produced the deepest 20cm radio survey
to date with 2050 sources and the basic radio properties of the faint $\mu$Jy
population. Paper II \citep{SDF2} details a complementary, deep
90cm survey and dependence of 20cm to 90cm spectral index
on radio flux density. Paper III \citep{SDF3} documents the WIYN 
spectroscopy of
sources in this field. This paper deals with
the first properties derived for the $\mu Jy$ population, namely the
photometric redshifts and inferred redshift distribution, and the
stellar population properties of the host galaxies. Throughout
this paper, we adopted the AB magnitude system and WMAP cosmology ($\Omega_{M}$= 0.27,
$\Omega_\Lambda$= 0.73, H$_{0}$= 71 km s$^{-1}$ Mpc$^{-1}$ ) unless otherwise
stated.

\section{Data}

This work is based on optical U,g,r,i,z, near--infrared (NIR) J,H,K,
IRAC $3.6 \mu$,$4.5 \mu$ and GALEX near--UV images of a patch $0.6
\times 0.6$ square degrees wide in the SWIRE Lockman Hole field,
hereafter the Deep Swire Field (DSF). This patch is approximately
centered on the region covered by deep VLA imaging ( 10$^h$46$^m$00$^s$, 
59\arcdeg01\arcmin00\arcsec ). An extensive spectroscopic campaign secured
spectroscopic redshifts for several hundreds objects, as detailed in
paper III.

Optical U, g, r images were obtained in 2002 (g,r), and 2004 (U) at
the Kitt Peak National Observatory (KPNO) Mayall 4m telescope. A
detailed description of these images, including data acquisition and
processing, can be found in \citet{polletta2006}.  A deep i-band image
was obtained from the Canada France Hawaii Telescope (CFHT) MegaCam
Science Archive. The data were acquired in 2005 during the observing
run 2005BH99. The stacked MegaCam image has been produced by the
MegaPipe pipeline at the CADC\footnote{For a detailed description of
the MegaPipe processing see
http://www2.cadc-ccda.hia-iha.nrc-cnrc.gc.ca/megapipe/index.html}.
Medium deep K band imaging, covering about 90\% of our field, has been
downloaded from the UK Infrared Telescope (UKIRT) Infrared Deep Sky
Survey (UKIDSS, \citet{ukidss}) science
archive\footnote{http://www.ukidss.org/archive/archive.html}. UKIDSS
uses the UKIRT Wide Field Camera (WFCAM, \citet{wfcam}) and a
photometric system described in \citet{ukidssphotsys}. The pipeline
processing and science archive are described in
\citet{wfcamhambly2008} and Irwin et al., in
preparation. We have used data from the DR2
data release, which is described in \citet{ukidssdr2}.  IRAC $3.6 \mu$
and $4.5 \mu$ images are part of the Spitzer Wide-area InfraRed
Extragalactic Legacy Survey (SWIRE,
\citet{lonsdale2003,lonsdale2004}).  GALEX NUV deep imaging has been
acquired on the DSF as part of the Deep Imaging Survey (DIS). Two
contiguous GALEX NUV pointings overlap on the DSF field. Deep stacks
have been made publicly available in early 2008 with the GALEX Release
4 (GR4)\footnote{http://galex.stsci.edu/GR4/}. In order to produce a
single image covering the DSF, the two GR4 images have been coadded
using the Swarp software \citep{swarp}.

Finally, we used proprietary and/or still unpublished z, J and H data.
Imaging in z-band was obtained with the MOSAIC camera on the 4m
telescope at KPNO on the nights of April 2--5, 2005. Ten hours of
on-sky integration was obtained for a pattern of pointings which
produced an image $48\arcmin\times48\arcmin$ in size centered on the
field. The image was reduced with the standard IRAF MOSAIC package.
Imaging in J and H band was obtained with WFCAM on UKIRT on the nights
of April 6--9, 2007. A total of eight hours on-sky was obtained for
each band to construct a mosaic image covering $54\arcmin
\times54\arcmin$ centered on the field. The data were pre--processed
with the UKIRT summit pipeline and then shipped to Cambridge for
further processing, including removal of instrumental signature, sky
subtraction, and stacking of microstepped images\footnote{see
http://casu.ast.cam.ac.uk/surveys-projects/wfcam/technical}.  The
Cambridge processed images were then mosaicked with the SIMPLE Imaging
and Mosaicking Pipeline\footnote{see http://www.asiaa.sinica.edu.tw/$\sim$whwang/idl/SIMPLE}.

\section{Catalogs}
\label{sec:catalogs}
Catalogs were generated with Sextractor \citep{sextractor} in
``dual--image'' mode. 
Three detection images were used, hereafter referred to as
``optical'' (g+i), ``NIR'' (J+H+K), and ``IRAC'' ($3.6\mu$m+$4.5
\mu$m). Even though the best strategy for building a detection image
would be to use all images at the same time \citep[e.g.,][]{szalaychi2image}, the significant difference in quality and
resolution among our images suggest that we can build a better
detection image by using only the best quality images we have (in
terms of resolution, artifacts, bad areas due to bright objects). For this
reason, the primary detection image was built from the g and i images
alone. In order to include in our analysis also redder objects, we
then considered the NIR and IRAC detection images as well.

All optical and NIR images were convolved with a gaussian kernel in
order to match the seeing of the worse optical image, namely the U
band with a seeing of $\simeq 1.3\arcsec$.  In order to measure photometry
in SExtractor dual--image mode, all images were registered on the
optical detection image pixels, with the IRAF tasks {\small GEOMAP}
and {\small GEOTRAN}. The accuracy of the pixel registering, as
measured with bright point--like sources, goes from 0.2-0.3 pixel
typical for the optical and NIR images, to $\approx$ 0.5 pixel for the
IRAC images (corresponding to less than 0.1 native IRAC pixel).

 Photometry was extracted in circular apertures of $0.75\arcsec$ and $1\arcsec$
radius. Since optical and NIR images were all smoothed at the same
resolution, their photometry needs no correction for different FWHM of
the original images. This is not the case for the GALEX and IRAC
images, which have a significantly worse resolution. In order to avoid
greatly degrading the resolution of the optical/NIR images, GALEX
and IRAC photometry was corrected separately for the effect of the
larger PSF, by means of aperture corrections based on point--like
source profiles matching the individual GALEX/IRAC resolution to the
optical/NIR matched resolution of $\simeq 1.3\arcsec$.

Mini-background maps produced by SExtractor were used to automatically
identify bad areas in each of the three detection images, i.e. areas
affected by large bright star halos, spikes and other kind of
artifacts, image corners poorly exposed, and generally all areas whose
background is not uniform with the rest of the image. While these
areas are discarded when dealing with statistical studies of the
optical/NIR/IR galaxy populations, we still consider them for the study
of the radio--selected galaxy sample, yet flagging the objects falling
in these areas for a subsequent visual follow-up. Furthermore,
all the single--band images were similarly analyzed in order to flag
areas where photometric quality cannot be considered uniform with the
rest of the image. Photometry in these areas was then handled with
particular care, as explained later. 

The accuracy of the aperture photometry measured on these images {\it
in dual--image mode} was tested with simulations, by adding to the
images point--like sources and attempting to recover them {\it given
their position is known}. It is important to stress that the image
quality/depth measured in this way doesn't take into account the
detection problem of faint sources in the single image, since we use
the detections coming from one of the three detection images described
above. Artificial sources with a gaussian profile of $1.35\arcsec$ for the
optical and NIR images, and $1.8\arcsec$ for the IRAC images were added
with the IRAF task {\small MKOBJECTS}. Assuming their positions were
known, their photometry was extracted in dual--image mode as for the
real objects, by creating bright enough objects at the same position
in the detection image. Again we note that this test only aims at
checking photometric accuracy of detected sources, independent of
detection issues: detection is assumed to be made on a higher S/N
detection image, possibly in a very different passband. Nonetheless,
these simulations include the effect of bad areas in both the
detection and photometry images, which affect the detection (in the
detection image) or the flux measurement (in the photometry image).
The retrieved aperture magnitudes were then compared with the input
(total) magnitudes (by applying the appropriate aperture corrections).
From these simulations, we estimated for each image the following
quantities, as a function of input magnitude: i) the percentage of
input objects for which SExtractor was able to measure a magnitude, ii) the
median difference $\Delta$mag between input and output magnitudes, and
iii) the 16$^{th}$-84$^{th}$ percentiles of the $\Delta$mag
distribution.

We used these quantities to determine, for each image, the magnitude
range where the measured flux can be considered meaningful, and a
realistic error on such flux. In particular, we discarded all
measurements fainter than the magnitude mag$_{cut}$ defined as the
faintest magnitude where more than 90\% of the input objects has a
measured flux, the median $\Delta$mag is less than 0.2 mag, and both
the 16$^{th}$ and 84$^{th}$ percentiles of the $\Delta$mag
distribution are within one magnitude from the input magnitude. While
these criteria allow to retain the advantages of the dual-image
extraction in measuring faint fluxes, they also allow to define for
each passband a limit beyond which such fluxes are no longer deemed
meaningful measurements, and thus are treated as drop-outs. From the
error curve derived from these simulations we also determine a
magnitude mag$_{10}$ where simulated objects have their flux measured
with a typical error of less than 0.1 mag. The adopted values of
mag$_{cut}$ and mag$_{10}$ are given in Table \ref{tab:magcutcompl}.

\section{The radio--selected sample}

The sample of radio sources used here is described in detail in
paper I. The DSF was observed with the VLA in
the A, B, C, and D configurations for a total of almost 140
hours on-source. The final image has a resolution of ~1.6\arcsec,
with a typical rms at the center of the image of 2.7 $\mu$Jy ( see paper I
for details).

\begin{figure}
\centering
\includegraphics[width=.49\textwidth]{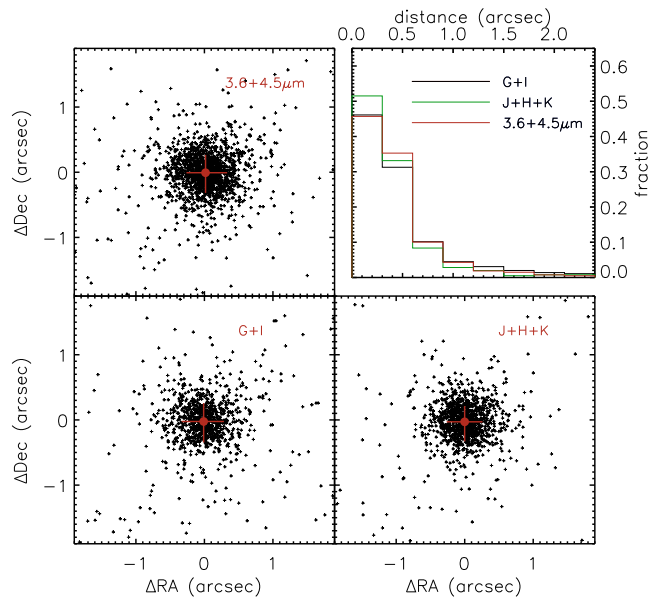}
\caption{$\Delta$RA vs $\Delta$Dec for radio--optical/NIR/IRAC matching
IDs (upper--left and lower panels). All matchings are shown for each
catalog within 2\arcsec. In each panel the red points show the median and the
errorbars show the $18^{th}$/$84^{th}$ percentiles of the $\Delta$RA
and $\Delta$Dec distributions. The upper--right panel shows the
histogram of the distances between radio IDs and matched counterparts,
independently for each of the three catalogs.
\label{fig:matdist}}
\end{figure}

The sky region included in our detection images contains 1930 sources
out of the 2055 of the original paper I radio catalog.  All three
(optical/NIR/IRAC selected) catalogs described in section
\ref{sec:catalogs} were used to identify a (optical/NIR/IR)
counterpart for the radio sources in the catalog.  The radio and
optical/NIR/IRAC WCS were first matched by correcting for the mean
offsets in $\Delta$RA$_{optical-radio}$ and
$\Delta$Dec$_{optical-radio}$. The scatter in
$\Delta$RA$_{optical-radio}$ and $\Delta$Dec$_{optical-radio}$ is
$\simeq$0.3\arcsec (see figure \ref{fig:matdist}). Then, for each
radio source the three catalogs were searched for a match within
$1\arcsec$ (with an order of preference 1)~optical, 2)~NIR, and
3)~IRAC. Most of the radio sample was matched with an optical
counterpart: out of the whole sample of matched counterparts, the
(optical unmatched) NIR counterparts and (optical and NIR unmatched)
IRAC counterparts contribute for about 12\% and 9\%,
respectively. However, about 20\% of these IRAC counterparts and 40\%
of these NIR counterparts are located in ``bad areas'' of the optical
detection image, where detection was hampered by bright sources or
artifacts.  Also, when restricting to the redshift range $0.3<z<1.3$
which will be the main focus in the following, IRAC selected
counterparts contribute for just 1\%, and NIR selected counterparts
for 7\% (half of which in bad areas of the optical detection image).

 While for most of the sources a counterpart closer than
1\arcsec~ was found, for almost 6\% of the 1930 sources it was not
possible to identify a counterpart within $1\arcsec$. Half of these
could be matched with a counterpart increasing the matching radius up
to $1.5\arcsec$. The other half of the sources without counterpart
were automatically and visually inspected: some of these sources are
indeed found in image areas affected by bright objects
halos/artifacts, some others are blended with another bright object --
clearly in such cases the possible counterpart may likely go
undetected. However, for 13 (out of 59) unmatched sources it seems
that there is actually no counterpart in our images (and no apparent
issues in the images which could hamper its detection).

In conclusion, 94\% of the radio sources contained in our detection
region were assigned a reliable counterpart (within $1\arcsec$), while
3\% were assigned a counterpart between $1\arcsec$ and $1.5\arcsec$ and 3\% were
not assigned any counterpart.

Since the resolution of the radio image is similar to that of the
optical/NIR images, we did not apply the likelihood ratio technique
\citep{richter1975} which is commonly used to evaluate the reliability
of each identification based on the source-counterpart distance and
counterpart magnitude. However, we used the distance and magnitude
distribution of our identified counterparts to evaluate the
contamination of our catalog from false counterparts, estimating the
probability of false association by randomly shifting the radio source
coordinates and repeating the association process. Given the
characteristics of most of our associations (75\% (97\%) are at a
distance of less than 0.5''(1'') from the source, 70\% brighter than
i=25 within an aperture of 1.5'') the overall contamination of the
whole catalog is estimated to be lower than 2\%. The contamination
from false counterparts of the $0.3<z<1.3$ sample which will be used
in most part of this work is estimated to be  negligible ($<1$\%).

Out of the sources with an assigned counterpart, 76\% were detected
(i.e. have a measured aperture magnitude brighter than mag$_{cut}$) in
the U band (and up to 80\% including objects possibly undetected
because in flagged areas of the U band image), 83\% (up to 89\%) in
the g band, 80\% (89\%) in the r band, 74\% (91\%) in the i band, 75\%
(92\%) in the z band, 85\% (91\%) in the J band, 90\% (95\%) in the H
band, 81\% (87\%) in the K band, 88\% (100\%) in the IRAC
3.6$\mu$ band, and 93\% (100\%) in the IRAC 4.5$\mu$ band. We note
that we give these numbers as an indication of the typical spectral
energy distributions (SEDs) of the
radio--selected sample: the specific numbers depend on the different
depth and overall quality in terms of bad areas of the different
images.

\section{Photometric redshifts}

\subsection{Determination of photometric redshifts from
 multi--wavelength NUV to mid--IR photometry}
\label{photozdet}

The optical, NIR and IRAC selected multi--wavelength catalogs
described above were used to estimate photometric redshifts
(photo--zs) by means of comparison with a library of galaxy SED
templates covering a range of star-formation histories, ages and dust
content. A set of 33 templates were used, spanning from a classical
local elliptical to several star forming galaxies to a QSO dominated
template, and all covering a rest-frame spectral window
[1000--70000]$\AA$, thus ensuring an adequate cross-correlation with
the available photometric coverage. Beside local galaxy templates
\citep[e.g.,][]{cww,mannuccitemplates,kinney1996}, a set of
semi--empirical templates based on observations plus fitted model
SEDs \citep{maraston1998,bc03} of $\approx300$ galaxies in the FORS
Deep Field \citep{heidt2003} and Hubble Deep Field
\citep{hdfwilliams1996}, were included to better represent objects to
higher redshifts.

Here we describe briefly the method used to estimate photo--zs, and we
refer to \citet{bender2001}, \citet{gabasch2004} and
\citet{brimioulle2008} for a more detailed description of this method,
as well as of the construction of the templates.  The aperture
PSF-matched photometry of each object in the available (up to 11)
passbands was compared with the templates, calculating a redshift
probability function over the range $0<z<10$ (in steps of 0.02) for
all SEDs. This is done assuming some priors: a different prior on the
redshift distribution is assumed for different types of templates
(corresponding to younger or older SEDs, that is e.g. an old local
elliptical template is assumed to be increasingly unlikely at higher
redshifts, while templates corresponding to young stellar populations
or QSOs are assumed to have a basically flat likelihood across all
redshifts explored). Furthermore, a weak, broad prior on the absolute
optical and NIR magnitude lowers the probability to have magnitudes
brighter than -25 and fainter than -13.  Ly-$\alpha$ forest depletion
of galaxy templates is implemented according to \citet{madau1995}.
The ``best--fit'' photo--z $z_{phot}$ is chosen as the
redshift maximizing the probability among all templates, and an error
on $z_{phot}$ is defined as
$e_{zphot}=[\Sigma_{ij}(z_{i}-z_{phot})^{2} P_{ij}]^{1/2}$, with
$z_{i}$ the considered redshift steps in $0<z<10$, and $P_{ij}$ the
contribution of the j-th template to the total probability function at
redshift $z_{i}$.

Systematic offsets between the measured and predicted colors as a
function of redshift, which may be due to several reasons as for
instance uncertainties in the estimated zero--point, but also possibly
uncertainties in the filter response curves or even systematics in the
templates, were estimated using almost 500 spectroscopic redshifts
available in our field (paper III, private communication from G. Smith
2007, plus some few more published redshifts available from
NED\footnote{http://nedwww.ipac.caltech.edu/}).  Therefore, the
zero--point for each of the passbands was corrected by a factor which
minimizes the systematic shift between observed and predicted color
for well--fitted spectroscopic galaxies.

Also stellar templates \citep{pickles} were fitted to all
sources. Comparing stellar and galaxy $\chi^2$ for objects with a
point--like morphology, we checked that objects brighter than
I$\simeq$23.7 with a best--fit stellar $\chi^2$ lower than the
best--fit galaxy $\chi^2$ could be classified as stars. The number
counts of such selected ``stars'' are in good agreement with
predictions of the \citet{robin2003} Galaxy model
\footnote{www.obs-besancon.fr/model/}. Fainter than
I$\simeq$23.7, the classification was found less reliable.
None of the sources in the radio sample was classified as star.

When fitting an object's photometry, besides excluding from the fit
all magnitudes deemed unreliable (including magnitudes of objects in
flagged areas), all magnitudes fainter than mag$_{cut}$ (as defined
above) were considered as drop-outs, and a few photo-z determinations
with different ways of dealing with drop-outs were compared, in order
to test the stability of the derived photo-z. Obviously, the estimated
photo-z is most unstable for those objects with a very high number
(e.g. $>6$) of drop outs (or combination of drop outs and flagged
magnitudes). Only a minor fraction of the objects is concerned, and
the effect on the global sample, and in particular at redshift lower
than 2, can be considered negligible. Nonetheless, the outcome of this
test was used together with the $\chi^2$ of the best fit, the error on
the estimated photo-z, and on the total number of actually measured
magnitudes used (e.g. not upper limits or flagged magnitudes), to
evaluate the quality of the estimated photo-z for each source.

Only ``constrained photo-zs'', i.e. with at least 4 {\it measured}
magnitudes actually used (thus not including upper limits and flagged
magnitudes) and a discrepancy between different determinations of less
than 20\% in $\Delta z $/(1+$z$) were deemed reliable\footnote{A
quality flag QF was defined as follows: QF=AA: a photo--z is
determined in all different realizations, the total number of upper
limits plus flagged magnitudes is less than 5, the maximum difference
$\Delta$z/(1+z) among the different determinations is less than 20\%,
the estimated error on the photo--z is less than 0.2(1+z). QF=A: same
as QF=AA but the constrain on the estimated error is relaxed to
0.4(1+z). QF=B: same as QF=A but constrain on number of flagged
magnitudes plus upper limits is increased to less than 7. QF=C1: same
as QF=B but no constrain on the estimated error on photo--z.}$^,$\footnote{  Also,
based on the $\chi^2$ distribution of such selected photo--zs and on
the comparison with the spectroscopic sample (see below), all
photo--zs having at the same time both a $\chi^2>2$ (97$^{th}$
percentile of the $\chi^2$ distribution of the ``constrained photo--zs'') and a
relative error $e_{zphot}/z_{phot}$ larger than 75\% were discarded.}.
In the following we only use photo--zs deemed reliable based on these
criteria, unless otherwise noted.

\begin{figure}
\centering
\includegraphics[width=.49\textwidth]{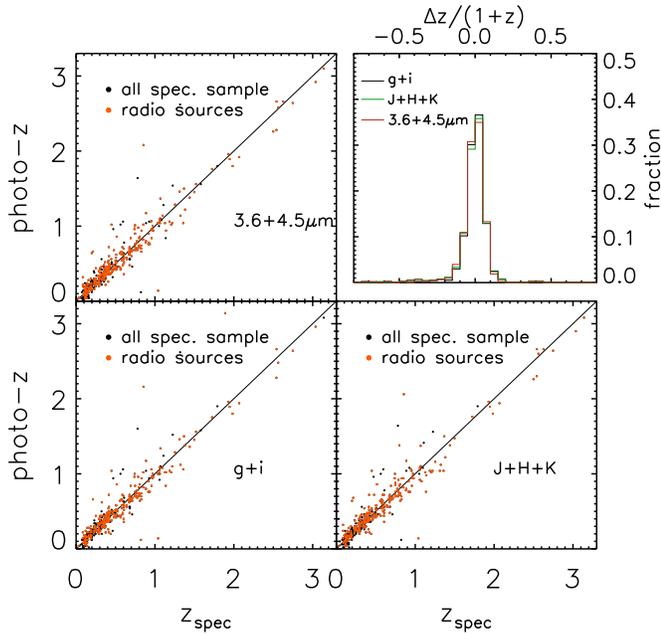}
\caption{The comparison of spectroscopic and photometric redshifts for
more than 400 sources in the field. In the upper--left and lower
panels the estimated photometric redshift is plotted against the
spectroscopic redshift, separately for the three optical, NIR and IRAC selected catalogs
(the numbers of matched sources are 402, 419 and 429
respectively). The solid line in each panel marks the bisector.
Orange symbols highlight radio sources, making up about 3/4 of the
spectroscopic sample. The histograms of $\Delta z$/($1+z$) for all three
catalogs are plotted in the upper--right panel.
\label{fig:zszpall}}
\end{figure}

~\\
\subsection{Accuracy of photometric redshifts}

The overall accuracy of our photo--zs is estimated by comparison with
the spectroscopic sample.  Figure \ref{fig:zszpall} shows the
comparison of the spectroscopic redshifts with the retained reliable
photo--zs for each of the photometric catalogs (optical, NIR and IRAC
selected). In each case, more than 400 objects have been used for the
comparison, deriving a median($\Delta z$/($1+z$))$<0.003$,
$\approx$3.5\% outliers (defined as $\Delta z$/($1+z$)$>0.2$), and an
accuracy of $\approx 0.05$ in $\Delta z$/($1+z$), as estimated either
from the NMAD estimator \citep{hoaglin1983,ilbert2009} or from the
16$^{th}$-84$^{th}$ percentiles of $\Delta z$/($1+z$).

\begin{figure}[b!]
\centering
\includegraphics[width=.483\textwidth]{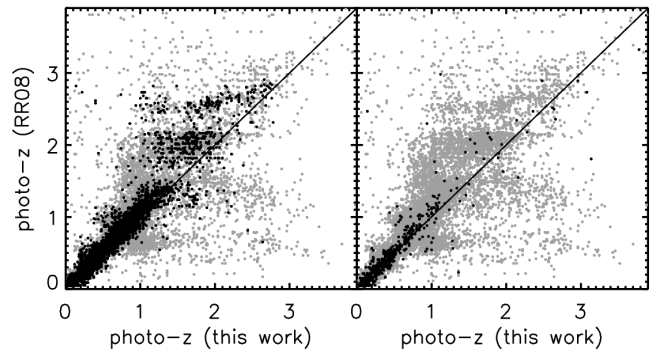}
\caption{The comparison of this work photo--zs with those derived by
RR08. Left panel: gray symbols show all data points while black
symbols show those which in the RR08 catalog have photometry in 6
passbands and mag r$<24$. Right panel: gray symbols show all data
points while black symbols show objects with available spectroscopic
redshift in either this work or RR08. See text for details.
\label{fig:zszpRR08}}
\end{figure}

While we have a sizable spectroscopic sample allowing us to assess the
accuracy of our photo--z determination, it may also be useful to
compare our results with independently determined photo--zs. While
such comparison has not the same strengths of the usual spectroscopic
vs photometric redshift comparison, it may help to overcome two of its
main weaknesses: spectroscopic samples are typically only a small
fraction of a galaxy sample, and are significantly biased toward
bright sources. In figure \ref{fig:zszpRR08} we show the correlation
between the photo--zs derived in this work and those derived by
\citet{RR08} (hereafter RR08). RR08 derived photo--zs for all the
SWIRE survey, including the Lockman Hole field used in this work. They
not only use a different code for estimating photo--zs , but also a
significantly different approach. While having a full optical+NIR
coverage in some of the fields, in the Lockman Hole they only use
photometry in U, g, r, i, 3.6 and 4.5$\mu$m passbands. They report for
the whole SWIRE survey a typical rms of
(z$_{phot}$-z$_{spec}$)/(1+z$_{spec}$), {\it excluding outliers},
slightly larger than 4\% for 6 passbands (which is the case of the
Lockman Hole) and r$_{Vega}<$24, and $\approx$ 4\% outliers.  In the
left panel of figure \ref{fig:zszpRR08} we thus highlight the
comparison for sources with r$_{Vega}<$24 and photometry available in
6 passbands in the RR08 catalog. By comparing our photo--zs with those
by RR08 for these objects, we find almost 6\% outliers, a median
(z$_{RR08}$-z$_{this work}$)/(1+z$_{this work}$) of $\approx -0.03$,
and a scatter $\approx$6\%, which is consistent with what expected
based on the claimed scatter of both works.  We note that their
performance, for all SWIRE fields, worsens for $z>1.5$ (see original
paper).  In fact, this worsening may be expected to be especially
significant for the Lockman Hole where no NIR photometry is used in
RR08, thus likely hampering photo--zs at $z>1-1.5$. In fact, as figure
\ref{fig:zszpRR08} shows, the comparison between our and RR08
photo--zs gets significantly worse beyond z$\approx 1.5$. As far as
our photo--zs are concerned, for $\approx 15$ objects with
$1.5<$z$_{spec}<3.2$ we have a median
(z$_{phot}$-z$_{spec}$)/(1+z$_{spec}$) $\lesssim 0.01$ and scatter
$\lesssim 0.05$, pretty similar to the figures for the whole sample,
but a formally higher number of outliers ($\lesssim$7\%, due to one
outlier out of $\approx$15 objects).  The right panel of figure
\ref{fig:zszpRR08} highlights instead the comparison of RR08 and our
photo--zs for sources with spectroscopic redshift available, either in
our catalogs or in RR08. For these sources, while finding a scatter
similar to that of the whole magnitude-- and number of passbands--
limited sample, we find a lower median ($\approx -0.01$), and $<5$\%
outliers.  We notice that the spectroscopic samples used in this work
and in RR08 have a significant overlap, thus this might affect the
comparison of the photometric redshifts results. Nonetheless, the
comparison of photometric redshifts derived with different photometry,
different SED libraries, and different methods, allows an indirect
evaluation of the photo--z performance on a much larger sample, and at
fainter magnitudes, than those allowed by the usual comparison with
spectroscopic redshifts. This comparison confirms, at least out to
redshift $\sim1.5$, the overall accuracy of our photo--zs as estimated
by comparison with our spectroscopic sample.

\begin{figure}
\centering
\includegraphics[width=.483\textwidth]{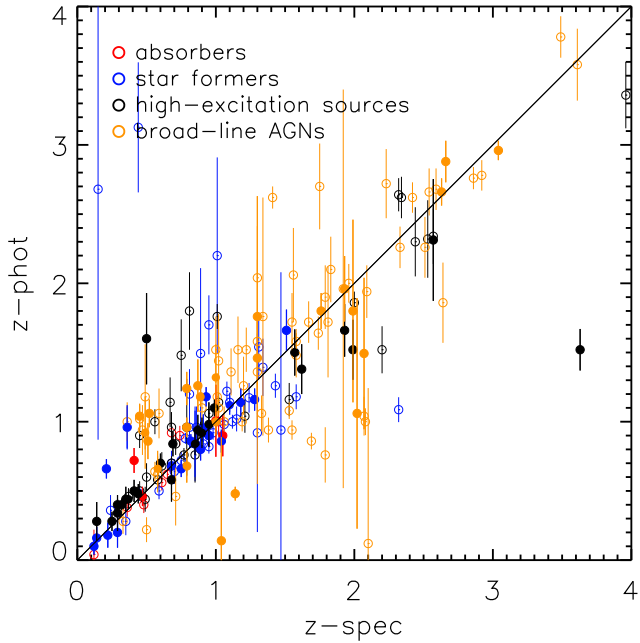}
\caption{The comparison of this work photo--zs with spectroscopic
  redshifts of 198 X--ray sources from the \citet{trouille2008}
  sample. Solid symbols refer to a subsample of 71 radio detections,
  while empty symbols do not belong to the radio sample. The solid
  line is the bisector. Color coding reflects the \citet{trouille2008}
  optical spectral classification as indicated (see text for details).
\label{fig:zztrouilleALL}}
\end{figure}

On the other hand, it is interesting to compare our photo--zs with a
spectroscopic sample of X--ray sources in this field published by
\citet{trouille2008}. In figure \ref{fig:zztrouilleALL} we compare our
photo--zs with \citep{trouille2008} spectroscopic redshifts for
$\approx$200 X--ray sources (out of which $\approx 70$ belong to our
radio sample) for which we have a deemed reliable
photo--z\footnote{Five more objects common to both the
\citet{trouille2008} and our sample were excluded from this comparison
because of possibly dubious spectroscopic redshift. One of these is a
z=0.35 object for which an available spectroscopic redshift is
consistent with our photo--z. The others all have redshifts z$\gtrsim
3$ in the Trouille et al. catalog, and thus might be expected to be
very faint or drop--out in the U band.  However they are all clearly
detected in our U band image, and in some cases in the GALEX NUV as
well.}. We divide and color--code the sample according to the
\citet{trouille2008} optical spectral classification, in absorbers,
star--formers, high-excitation sources and broad-line AGNs (see the
original paper for details). The fraction of such ``peculiar'' objects
in this comparison sample is quite relevant, with 40\% of the sample
being made of broad-line AGNs, and a further $\approx$ 30\% of
high-excitation sources.  As figure \ref{fig:zztrouilleALL} shows, our
photo--zs perform significantly worse for this spectroscopic sample,
as compared to our (or RR08) spectroscopic sample, with much larger
scatter and number of outliers. The overall statistics for
$\Delta$z/(1+z) for this sample is median$\leq -0.015$, NMAD scatter
$\approx$10\%, and $\approx$20\% of the objects having
$\Delta$z/(1+z)$>$20\%. As it is clear from the figure, the worse
results are obtained for the broad-line AGN sub--sample (NMAD scatter
$\approx$17\%, $\approx$30\% of the sample with
$\Delta$z/(1+z)$>$20\%).  The poor agreement obtained from this
comparison is not unexpected, due to the contamination of the
broad--band photometry with a strong AGN contribution \citep[see
e.g.][and references therein]{polletta2007,salvato2009}. We also
note that our photo--zs performance appears to be similar or better
than that achieved by \citet{trouille2008}, even though when
estimating photo--zs they include templates built from their
spectroscopically confirmed broad-line AGNs, while our template set
was not specifically tailored toward AGN--dominated SEDs. Furthermore,
we note that many (not all though) of the outliers have a large error
associated with the estimated photo--z.

\subsection{Photometric redshifts for the radio sample}

An estimated reliable photo--z is determined for 1610 radio sources
(86\% of the identified counterparts and 83\% of the whole radio
sample included in the detection image), while for the rest either no
photo--z could be estimated, or in most cases it was deemed not
reliable according to the criteria defined above.

In figure \ref{fig:zszpall}, the $\approx 300$ radio sources making up
about 3/4 of the spectroscopic sample are highlighted with orange
symbols. While one might expect a worse performance of the photo--zs
for the radio sample, because of the presence of the significant AGN
population among radio sources, the statistics for the radio sample
alone are quite similar, and only slightly worse, than those obtained
using the whole spectroscopic sample: median $\Delta
z$/($1+z$)=0.0008, scatter $\approx$5.5\%, and 4\% outliers (however
we note again that a large fraction of the whole spectroscopic sample
is made of radio sources). In figure \ref{fig:zztrouilleALL} we have
already shown (solid symbols) how our radio sample photo--zs perform
for the \citet{trouille2008} X--ray selected sources. In the
following, we include the spectroscopic redshifts from
\citet{trouille2008} in our analysis of the radio sample.

\begin{figure}
\centering
\includegraphics[width=.49\textwidth]{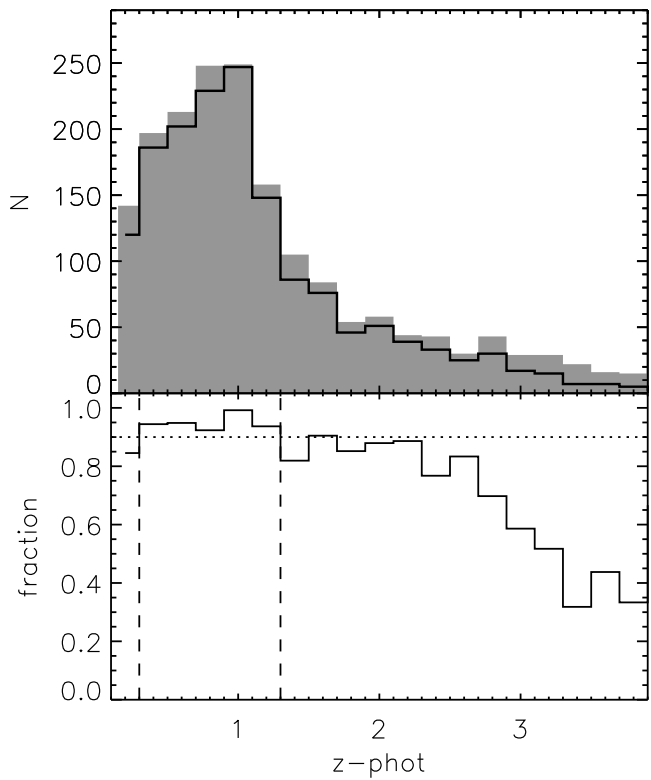}
\caption{The redshift completeness of the radio sample. {\bf Upper
    panel:} The shaded histogram shows one realization of the redshift
    distribution of the whole sample of identified counterparts, while
    the solid line shows the histogram of used redshifts
    (spectroscopic redshifts or reliable photo--zs). {\bf Lower
    panel:} The fraction of available redshifts (spectroscopic
    redshifts or reliable photo--zs) as a function of
    redshift: the sample is assumed to be more than 90\% complete in
    the redshift range $0.3<z<1.3$.  \label{fig:zphotselrange}}
\end{figure}

In figure \ref{fig:zphotselrange} we show the redshift completeness of
our sample (including spectroscopic and reliable photo--zs) as a
function of redshift\footnote{In the following, 8 objects with QF=C1,
unusually low $\chi^2$ and very high estimated photo--z error were
excluded from the reliable photo--z sample.}. Our criteria for
selecting a reliable photo--z naturally disfavor higher redshift
sources.  Since of course we do not know the redshift of all the
sources, we can do just an approximate estimate: we assumed that {\it
all} our photo--zs were broadly correct, even those we decided were
unreliable, and used them as a reference to estimate the redshift
completeness of the sample. By comparison with our different photo--z
catalogs based on different settings (of which figure
\ref{fig:zphotselrange} is one example), we estimated that in the
redshift range 0.3--1.3 our redshift completeness should always be
higher than 90\%, and the $0.3<z<1.3$ sample is overall complete at a
95\% level. In other words, if our photo--zs are broadly correct so
that the total number of objects in the redshift range $0.3<z<1.3$ is
right, we estimate an overall redshift completeness {\it in the
$0.3<z<1.3$ redshift range} of more than 95\% of the identified
counterparts, and more than 90\% of the whole radio sample within the
detection image\footnote{{\it Within the detection image} here means
within the boundaries of the detection images and excluding regions
within the boundaries where all three detection images were masked.}
assuming that all radio sources without an identified counterpart
are in this redshift range. This is definitely a conservative
assumption since it is likely that many of the unidentified
counterparts are very faint galaxies probably at higher redshift.

\begin{figure}[b]
\centering
\includegraphics[width=.485\textwidth]{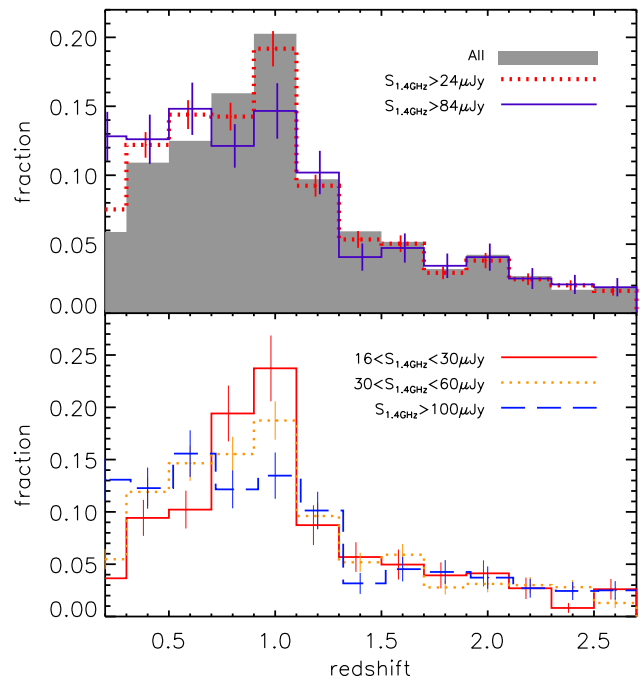}
\caption{ {\bf Top panel:} the gray shaded histogram shows the
redshift distribution in the range $0.1<z<2.7$ based on the sample of
radio sources with identified counterpart and reliable redshift
estimate (see text).  The red (dotted) and purple (continuous) lines
show the redshift distribution of flux--limited samples with radio
flux density above 24$\mu$Jy and 84$\mu$Jy, respectively. All
distributions are sky-coverage corrected (see text).  All histograms
are normalized to the total number of objects in each sample. {\bf
Bottom panel:} The red (continuous), yellow (dotted) and blue (dashed)
lines show the redshift distributions of sub--samples with radio flux
densities in the range [16--30], [30--60], and $>100\mu$Jy,
respectively. In both panels, all distributions are sky-coverage
corrected (see text), and are normalized to the total number of
objects in each sample. Errorbars show Poissonian errors.
\label{fig:zdist}}
\end{figure}

\subsection{Redshift distribution of the radio sample}
\label{sec:radiophotoz}

In figure \ref{fig:zdist} we show the redshift distribution of radio
sources based on the radio (sub--)sample for which either a reliable
photo--z or a spectroscopic redshift is available (gray shaded
histogram).  The redshift distribution is shown up to redshift
$z\sim2.7$, beyond which the redshift completeness is expected to drop
below 80\%\footnote{We also remind the reader that the sample shown here includes
a negligible fraction (less than $2\%$) of counterparts matched within
1.5\arcsec.}. The redshift distribution plotted is not corrected for
the estimated redshift incompleteness as estimated above (e.g. figure
\ref{fig:zphotselrange}).

\begin{figure}
\centering
\includegraphics[width=.49\textwidth]{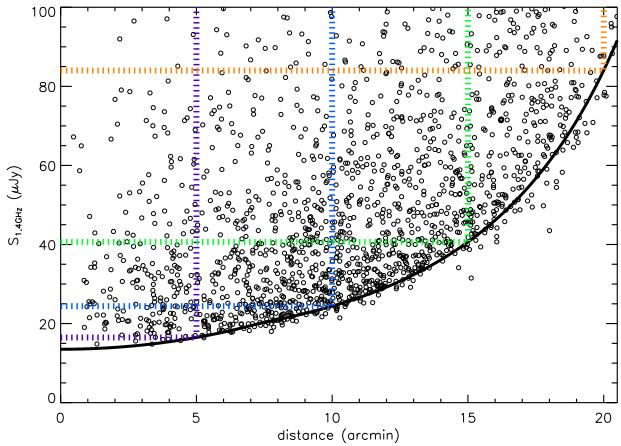}
\caption{The solid line shows the 5--$\sigma$ limiting flux density as
  a function of distance from the center of the VLA field, for a
  source size of 1.2'' (see text for details). Empty symbols show the
  integrated flux density vs. distance from the field center for
  actual detections. Thick dotted lines of different colors show four
  examples of flux-limited sub--samples selected in uniformly covered
  portions of the image ($R<5\arcmin$ and $S_{1.4GHz} > 16\mu$Jy,
  $R<10\arcmin$ and $S_{1.4GHz} > 24\mu$Jy, $R<15\arcmin$ and
  $S_{1.4GHz} > 41\mu$Jy, and $R<20\arcmin$ and $S_{1.4GHz} > 84\mu$Jy).
  \label{fig:rmscorr}}
\end{figure}

Since the DSF radio image is obtained with a single VLA pointing,
the rms is not constant over the field, but increases with the
distance from the field center. For this reason, the whole radio
sample does not have a single flux density limit and this needs to be
taken into account. In the following, wherever it is needed we will
correct the biased nature of the whole sample by taking into account
the non--uniform rms. While it is virtually impossible to make a
perfect correction, due to the unknown size distribution of the radio
sources, for our purpose we will calculate the appropriate correction
for all resolved sources for which a reliable size could be estimated,
and will assume that all remaining sources have a typical source size
of 1.2\arcsec (see paper I).  Taking into account the bandwidth and
time smearing, as well as the primary beam correction, and the
resolution of the image where the detection was performed, we can
estimate how the rms changes across the image, and thus the
5--$\sigma$ limiting flux density\footnote{The radio sample used in
this work only includes detections with $S/N>5$.}  at each distance
from the field center (see paper I for more details). In figure
\ref{fig:rmscorr} we show, as an example, the estimated 5--$\sigma$
limiting flux density as a function of distance from the field center
for a source of size$\sim$1.2''. Based on this estimated 5--$\sigma$
limiting flux density as a function of distance, and on the masked
areas in the optical/IR detection images, we calculated the area
$A_{i}$ over which each object could be observed and enter our
sample. The sky coverage is then used to weight each object (by
1/$A_{i}$) and thus estimate properties for a flux--limited sample. We
note that, as it can easily be seen from figure \ref{fig:rmscorr}, the
faintest sources in our sample can be observed over a very small
fraction of the survey ($<$10\%). One might thus be concerned that
inaccuracies in the sky coverage correction would significantly affect
the results. However, we note that all results presented in the
following would be unchanged if considering only flux-limited
sub--samples selected in uniformly covered portions of the image
(e.g., $R<5\arcmin$ and $S_{1.4GHz} > 16\mu$Jy, or $R<20\arcmin$ and
$S_{1.4GHz} > 84\mu$Jy,  see fig.\ref{fig:rmscorr}), while the sky coverage correction allows us
to make full use of our data set. In the following, we will explicitly
refer to flux--limited samples when using sky-coverage corrected
samples, in contrast with the whole, biased sample.

All distributions plotted in figure \ref{fig:zdist} are corrected
by the sky coverage. Beside the redshift distribution derived from the
whole sample, in the top panel of figure \ref{fig:zdist} we also show
the redshift distributions of two flux--limited samples with
$S_{1.4GHz} >$24 and $>$84 $\mu$Jy. These suggest that the redshift
distribution might depend of the flux density of the sample, and we
try to make this clearer in the lower panel of figure \ref{fig:zdist},
where we plot the redshift distributions of three sub--samples in
different flux density ranges, our faintest sources
(16.5$<S_{1.4GHz}<$30 $\mu$Jy), our typical about--median flux
population (30$<S_{1.4GHz}<$60 $\mu$Jy), and bright sources
($S_{1.4GHz}>$100 $\mu$Jy). The redshift distribution of the faintest
sources seems to be different from that of the bright ones, with a
sharper peak at $z\approx0.9$. A Kolmogorov-Smirnov test indeed
suggests that the two distributions are different at a high
significance level ($P \leq 0.005$)\footnote{We note that the
Kolmogorov-Smirnov results quoted here and in the following are
obtained from flux-limited sub--samples selected in uniformly covered
portions of the image, since the test cannot be meaningfully applied
to the sky--coverage corrected data}.

\section{Spectral energy distribution fitting}
\label{mmll}

We use SED fitting on the available multi--wavelength photometry in
order to estimate fundamental properties of the stellar populations
hosted in this radio source sample.  In the following we will consider
photo--zs just as spectroscopic redshifts, assuming that the galaxy is
at that redshift and ignoring any error on the photo--z. 
Different SED fits are used in the following. The first
characterization of each galaxy SED is given by the best--fit template
associated with the best--fit photo--z. As already said above, and as
will be described more in detail below, the templates span a range in
stellar population ages, including dust extinction. These templates
will then be used to classify galaxies according to the broad, average
properties of their stellar populations. In order to properly treat
galaxies with an available spectroscopic redshift, the fit was
recomputed for these objects assuming the spectroscopic redshift.

In order to classify SEDs based on a more ``parametric'' approach, and
also to investigate possible misinterpretations coming from the use of
non--evolving templates, we also performed SED fits using stellar
population synthesis models produced with the \citet{bc03} code. Star
formation histories (SFHs) were parametrized by simple exponentially
declining star formation rates, with a timescale $\tau$ ranging
between 0.1 and 20 Gyr, and age between 0.01 Gyr and the age of the
Universe at the object's redshift. Metallicity is fixed to solar and a
\citet{salpeter1955} initial mass function is adopted, with lower and
upper mass cutoffs of 0.1 and 100 M$_{\odot}$. A variable amount of
extinction by dust is also included, with $A_V$~$\in [0, 1.5]$. The
fitting procedure is described in full detail in
\citet{drory2004b,drory2005}. These results too will be used in the
following to roughly characterize the host stellar populations based
on the best--fit age/$\tau$ (meaning the age of the stellar
populations from the onset of the star formation divided by the
e--folding time of the exponentially declining SFH).

Finally, SED fitting is also used to estimate stellar masses. For this
purpose we use, as it is customary, a two-component model, adding to
the main smooth component (exponentially declining SFHs described
above) a secondary burst. This burst is modeled as a 100 Myr old
constant star formation rate episode. Metallicity and IMF of the burst
are the same assumed for the main component, however dust extinction
for the burst is allowed to reach higher values ($A_V$~$\in [0, 2]$).
We notice that since we fit aperture photometry, the stellar masses
obtained refer to the portion of the galaxy contained in the
aperture. We correct these masses to ``total stellar masses'' by means
of the ratio of ``total'' (FLUX\_AUTO) and aperture fluxes in the
detection images. The median correction applied for the radio sample
is a factor $\approx 2$, and 90\% of the sample is corrected by a
factor ranging between 1.5 and 4.  This is an approximation
neglecting any color gradients which may certainly exist in the
galaxy, in other words we assume that the stellar populations within
the measured aperture are representative of the stellar populations of
the whole galaxy.

Some of the main derived properties which are used in this work,
for objects with spectroscopic or reliable photo--z, are listed in
Table \ref{tab:bigtable} which is available in full on the on--line version.

\begin{figure}
\centering
\includegraphics[width=.485\textwidth]{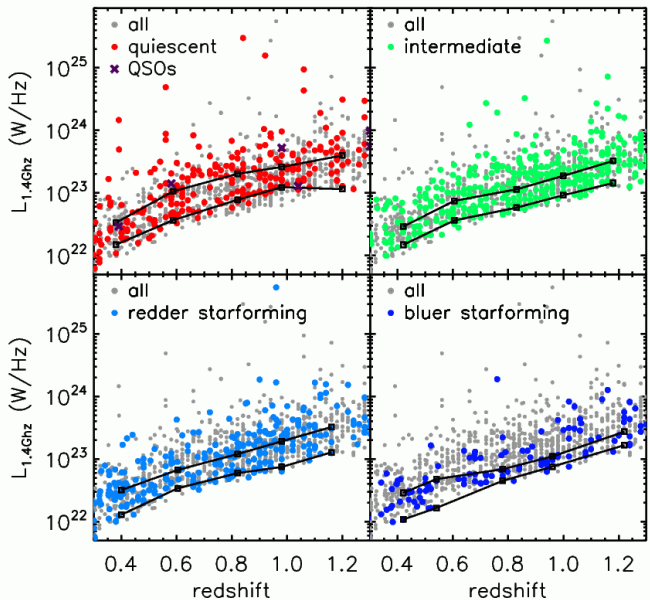}
\caption{The 1.4 GHz luminosity plotted against redshift for objects
with redshift between 0.3 and 1.3. Grey points in all panels show all
data points while colored symbols show objects whose photometry was
best--fitted by SEDs of different types, as indicated in each
panel. In each panel, connected small squares show the (sky coverage
corrected) interquartile range of the $L_{1.4GHz}$ distribution as a
function of redshift for the relevant SED--selected sub--sample.
\label{fig:zlumSEDs}}
\end{figure}

\section{SED properties of host galaxies: AGN activity and star formation}

In the following we compare the radio and optical/NIR properties of
this radio sample based on the SED fitting results described
above.

As shown in figure \ref{fig:zlumSEDs}, galaxies best--fitted by
different kind of templates (thus in principle galaxies with different
stellar populations) tend to occupy different locations in the radio
luminosity against redshift diagram. The simplest, and expected,
explanation is that different radio luminosities are associated to
different physical processes, namely star formation and AGN
activity. Figure \ref{fig:zlumSEDs} shows the radio luminosity against
redshift for classes selected based on the best--fit photo--z SED
template. The templates were divided in ``quiescent'' (including for
instance the elliptical template by \citet{cww} and the S0 and Sa
templates by \citealt{mannuccitemplates}), ``intermediate'' galaxies
with low star formation (including e.g. the Sb
\citet{mannuccitemplates} template), and ``star--forming'' templates
including all actively star--forming galaxies, plus a ``QSO'' class of
a few objects best--fitted by a QSO template.  As a reference, the
restframe U-B color\footnote{Restframe U-B colors, here and in the
following, are calculated in the \citet{buser1978} U and B3 filters.}
ranges for the three classes of quiescent, intermediate, and
star--forming objects are approximately 1.1--1.4, 0.9--1.1, 0.1--0.9,
respectively. Similarly, the U-V color ranges are approximately
1.9--2.2, 1.5--1.8, 0.1--1.3, while the break strengths at 4000$\AA$,
$D_{n}(4000)$ \footnote{We adopt the \citet{balogh1999} definition of
the $D_{n}(4000)$ index, that is the ratio of the average flux
densities in the narrow bands 4000--4100 $\AA$ and 3850--3950 $\AA$.},
are about 1.6--2.1, 1.5--1.7, 1--1.3. For reference, these
$D_{n}(4000)$ ranges may be compared for instance with typical
$D_{n}(4000)$ values for different kinds of stellar populations
measured in a local (SDSS) sample \citep[][their figure
7]{gallazzi2005}, and in a VVDS sample in the redshift range $0.45<z<1.2$
\citep[][their figure 4]{franzetti2007}. We also note that in figure
\ref{fig:zlumSEDs}, the ``star--forming'' class is further split in
two sub--classes of star--forming templates, ``redder SF'' and ``bluer
SF'', with restframe U-B ranging in 0.7--0.9 and 0.1--0.7, and
$D_{n}(4000)$ about $\approx$1.3, 1--1.2, respectively.

Due to the non--evolving nature of the templates used in the photo--z
determination, figure \ref{fig:zlumSEDs} only shows the stellar
populations status (i.e. actively star--forming, passively evolving,
etc.) {\it at the time of observations}, without any evolutionary link
between same--class objects at different cosmic epochs. In other
words, depending on the specific star formation history of each
galaxy, and on the overall evolution of galaxy stellar populations,
galaxies may (and will) change their class as time goes by. 

As figure \ref{fig:zlumSEDs} shows, at all redshifts the highest radio
luminosities in the probed range (e.g. L$_{1.4GHz}> 2\times10^{23}$
W/Hz at $z\sim0.5$ or L$_{1.4GHz}> 5\times10^{23}$ W/Hz at $z\sim1$)
are typical of low--starforming systems (i.e., galaxies classified as
intermediate or quiescent). Nonetheless, radio luminosities of such
low--starforming galaxies span all the range covered by this
survey. On the other hand, galaxies classified as actively
star--forming tend to avoid the highest radio luminosities, with just
few exceptions. The median radio luminosities of all sub--samples
are similar, with star forming galaxies typically showing a median
radio luminosity about 20\% lower than low-starforming systems
(quiescent and intermediate), over the redshift range probed. However,
the mean radio power of intermediate and star forming galaxies is
typically lower than that of quiescent galaxies (the mean L$_{1.4GHz}$
of bluer star--forming, redder star--forming, and intermediate
galaxies are 30\%, 40\%, and 50\% of the mean L$_{1.4GHz}$ of
quiescent classified sources, respectively).  The interquartile ranges
for L$_{1.4GHz}$ plotted in figure \ref{fig:zlumSEDs} also show how
the typical range and spread in radio power is different for different
SED classes, with the bluer star--forming galaxies mostly lying just
above the flux density limit of this survey.

Finally, figure \ref{fig:zlumSEDs} shows how a significant part of
this radio sample is made of objects classified as intermediate.  This is
likely due to the depth of this survey which allows us to go beyond
the AGN dominated population, but at the same time does not allow us
to reach the still fainter radio luminosities typical of the bulk
of normal star--forming galaxies. We note that, while the statistics
given above take into account the sky coverage, the points plotted in
figure \ref{fig:zlumSEDs} show the whole radio sample with
$0.3<z<1.3$, regardless of the different sensitivity at different
radii in the VLA image, therefore the different densities of data
points are not directly indicative of the actual relative fractions of
different kinds of objects, which are better addressed below.

\begin{figure}
\includegraphics[width=.485\textwidth]{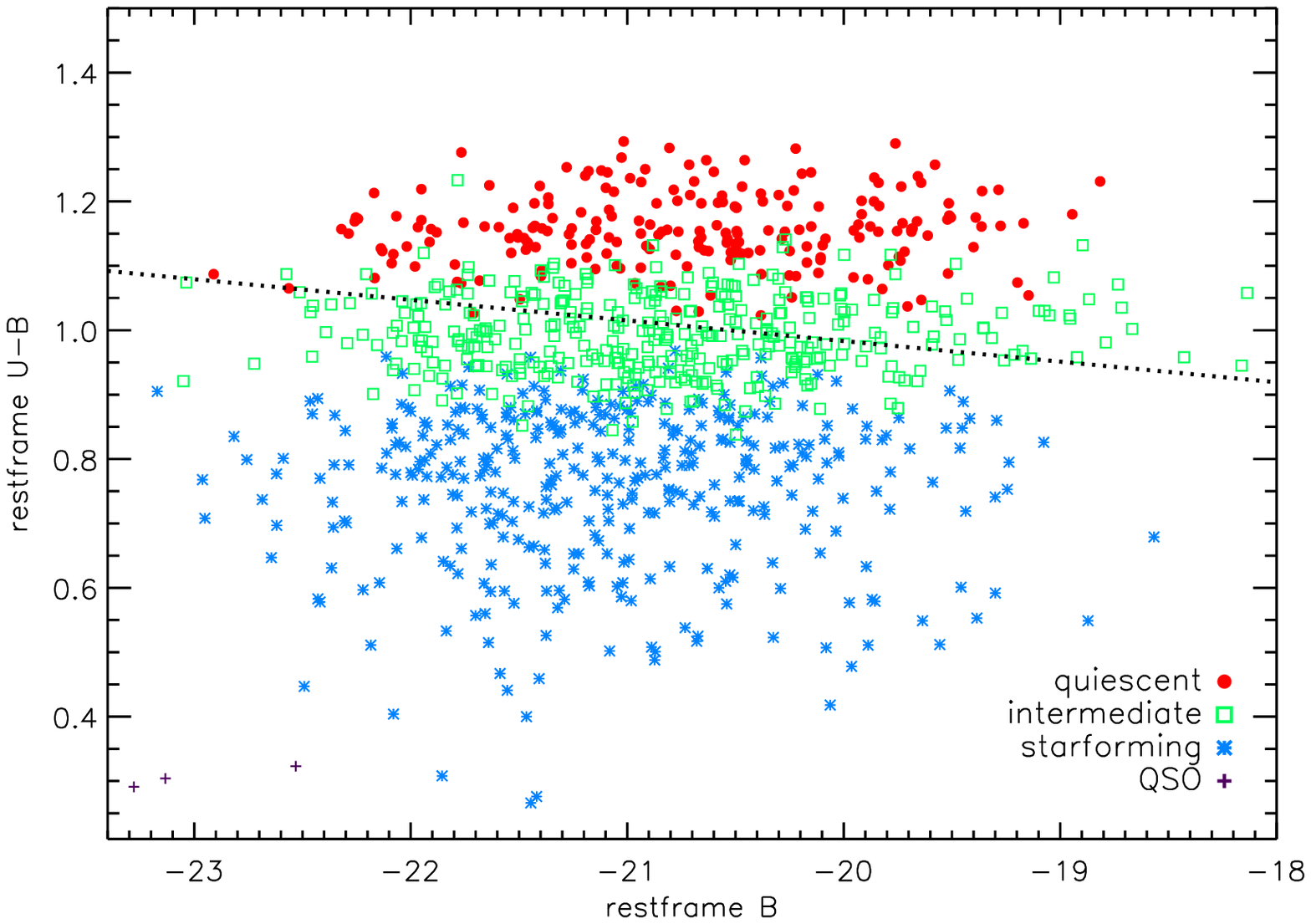}

\includegraphics[width=.485\textwidth]{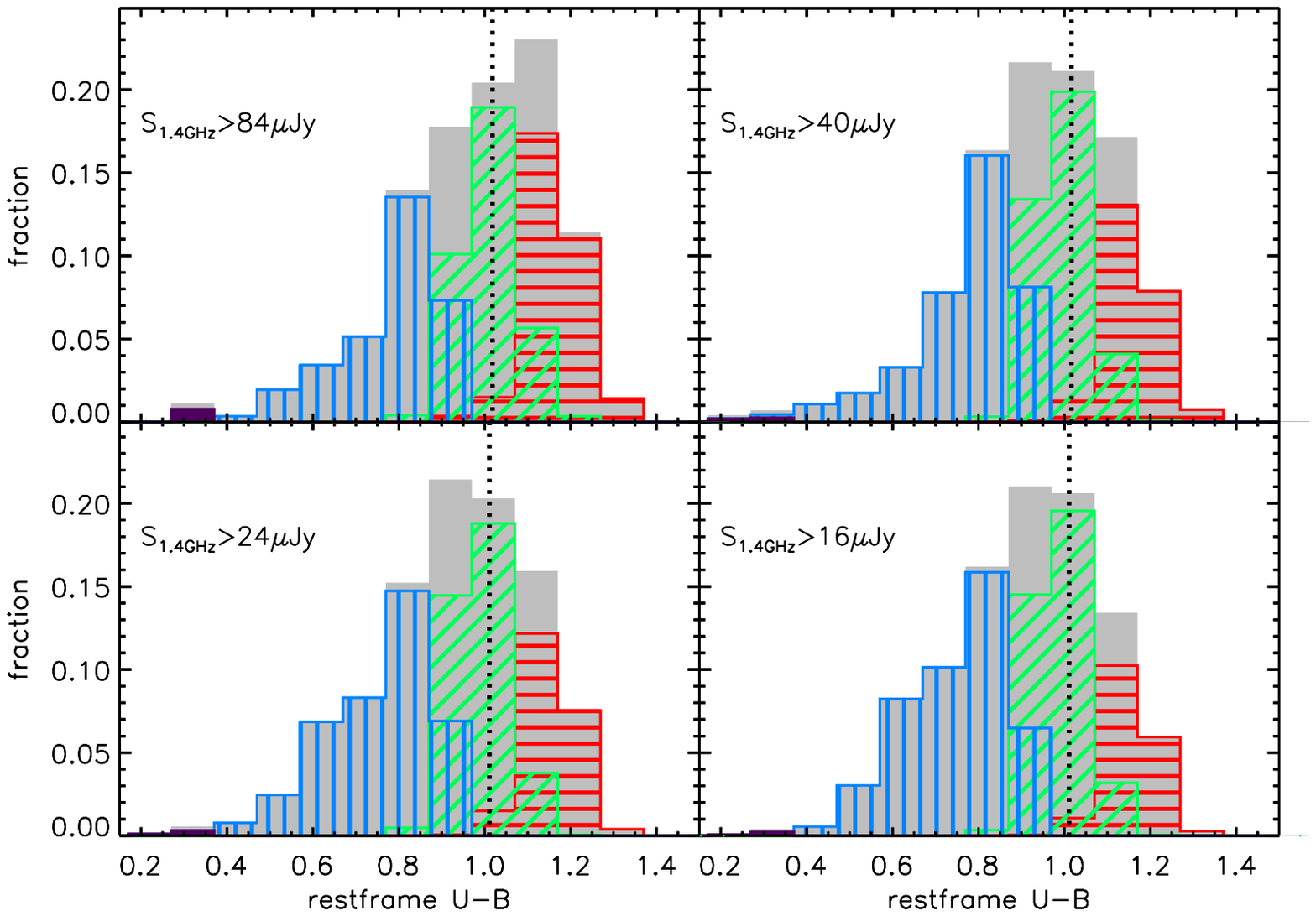}
\caption{{\bf Top panel:} The restframe U-B vs B color--magnitude
 diagram for objects with redshift $0.3 \leq z \leq 1.3$. All objects
 are included regardless of their radio flux density and position in
 the VLA image. Symbols are color coded according to best--fit SED
 type, as indicated (same broad classes as in figure
 \ref{fig:zlumSEDs}). The dotted line shows the division between red
 sequence and blue cloud from DEEP2 data as in
 \citet{willmer2006,cooper2007}. {\bf Bottom panels:} The distribution
 of restframe color U-B for the different SED classes, in four
 flux--limited sub--samples, as indicated (see text for details). The dotted lines mark
 the U-B color of the division line at the median B magnitude of each
 plotted sample. In each panel, histograms for the different SED
 classes are color--coded as in the upper figure (quiescent =
 horizontal, intermediate = diagonal and star--forming = vertical
 hatched), and the gray--shaded histogram shows the distribution of
 the whole subsample plotted in the panel.
\label{fig:U-BSEDs}}
\end{figure}

\subsection{The nature of faint radio sources: a significant intermediate population?}

In figure \ref{fig:U-BSEDs}, we show the restframe U-B vs B
color--magnitude diagram of the radio sample, divided as above in
quiescent, intermediate and star--forming.  In this figure, all
sources with redshift $0.3<z<1.3$ are plotted\footnote{However, 22
spectroscopic objects are not plotted because, due to a high number of
flagged magnitudes, they lack sufficient photometric information to
perform a reliable SED fit, even with known spectroscopic
redshift. This negligibly lowers the completeness of the sample
plotted, which is now 94\%.}, taking advantage of the small evolution
of the restframe U-B color in the redshift range probed, and thus just
one separation between red and blue galaxies was adopted at all
redshifts \citep[see e.g.][in the same redshift range studied
here]{willmer2006,cooper2007}.  The figure shows the well known
different locations in the color--magnitude diagram of different
galaxy types, with quiescent galaxies on the red--sequence and
star--forming galaxies in the blue--cloud. However, as compared to a
typical color--magnitude diagram for an optically selected galaxy
sample, it is evident that the so--called ``green valley'', often
thought to be populated by transitional objects which are shutting off
their star formation and migrating to the red sequence, is not an
underpopulated region in this diagram, and actually contains a
significant fraction of this radio--selected sample.

\begin{figure}[b]
\centering
\includegraphics[width=.49\textwidth]{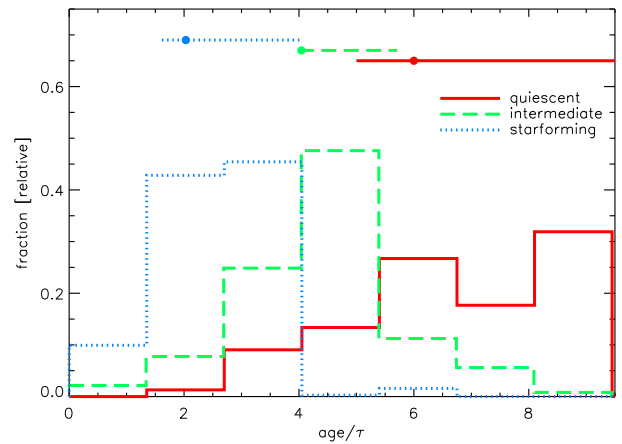}
\caption{The age/$\tau$ distribution for the sub--samples of
  quiescent, intermediate, and star--forming objects. The solid
  symbols and errorbars on top of the histograms show the median and
  16$^{th}$-84$^{th}$ percentiles of each distribution. Age/$\tau$
  values larger than 9.5 were set to 9.5 for the purpose of this plot.
\label{fig:agesutau}}
\end{figure}

However, the actual nature of this significant population of
apparently intermediate--type galaxies with low star--formation activity
is not necessarily clear. In fact, while it might be tempting to
consider the option that these are indeed ``composite'' or
``transition'' systems, meaning galaxies shutting off their star
formation because of (or linked to) an host AGN, on the other hand the
colors characteristic of an ``intermediate'' SED can also be produced
in other ways, the most obvious being a starburst affected by a very
high dust extinction which is not properly handled by our template
set, or possibly a red--sequence galaxy whose observed SED is severely
contaminated by AGN emission. In fact, in an attempt to look a bit
further into the dust attenuation issue of the intermediate galaxies,
we can compare our template--based classification with the parametric
one based on the single--component model SED fitting. As shown in
figure \ref{fig:agesutau}, there is a good overall correlation between
the age/$\tau$ of the best--fit model and the template--based
classification. However, if we look at the dust extinction which
according to the best--fit is affecting the model SED, we find that
10\% of the intermediate sample at $0.3<z<1.3$ was best--fitted by a
significantly extincted young stellar population (age/$\tau <4$, $A_V
>1$). For comparison, 30\% of the galaxies classified as star--forming
was best--fitted by such kind of model, and 3\% of the galaxies
classified as quiescent.  Even though we will not attempt to correct
colors based on the rough attenuation estimated by this simple SED
fit, the occurrence of apparently significantly extincted objects in
the intermediate sample may suggest a possibly relevant contamination
by reddened starbursts \citep[see also e.g.][]{cowie2008}.  The actual
nature of these intermediate sources will be further investigated in a
forthcoming paper, by exploiting X--ray and infrared data.

For the time being, to avoid misinterpretations, in the following we
will drop the labeling of the three classes as quiescent,
intermediate, and star--forming, which directly refers to the inferred
nature of the host stellar populations, and will adopt a more generic
and observationally--based ``red'', ``green'' and ``blue''.

It is worth noticing how the nature of the radio--galaxies host
stellar populations significantly depends on the flux densities
reached by the radio survey. In the lower panels of figure
\ref{fig:U-BSEDs} the sample plotted in the upper panel has been split
in order to draw more meaningful conclusions on the stellar
populations of flux--limited radio--selected sub--samples.  As it is
evident from figure \ref{fig:U-BSEDs}, and as it is expected (see
section~\ref{intro}), the stellar population properties of the host
galaxies, and thus likely also the process responsible of the radio
emission, are different in different radio flux density ranges. In the
shallowest sample considered ($S_{1.4GHz} > 84 \mu$Jy), the fractions
of blue, green, and red classified objects are roughly similar (33\%,
36\%, 30\%, respectively); in the $S_{1.4GHz} > 40 \mu$Jy and
$S_{1.4GHz} > 24 \mu$Jy samples they are about 40\%, 40\%, 20\% , and
eventually they become 45\%, 33\%, 22\% in the deepest radio sub--sample
($S_{1.4GHz} > 16 \mu$Jy).  This clearly suggests how the
contributions of the actively star--forming and red galaxies change with
the limiting flux density of the sample, with red galaxies increasing
their relevance in brighter samples and, viceversa, star--forming
galaxies becoming more important in fainter samples. For these
flux--limited samples, according to a Kolmogorov--Smirnov test, the
U-B color distributions of the radio--faintest and brightest
sub--samples are different at a $\simeq 98\%$ significance
level. However, the significance of the change in the colors of the
host galaxies at different radio flux densities is obviously more
evident when considering sources in {\it ranges} of radio flux density
instead of flux--limited samples, as we will do in the following.

We should notice that, since the lower panels of figure
\ref{fig:U-BSEDs}~ refer to a flux--limited sample, there are
different effects which come into play in realizing the different
distributions of the different--depth sub--samples, as for instance the
higher fraction of $z \simeq 1$ as compared to $z<0.5$ objects probed
at fainter flux densities (see figure \ref{fig:zdist}), and likely an
evolution with redshift of the 20cm luminosity threshold between star
formation--dominated and AGN--dominated radio samples (see figure
\ref{fig:zlumSEDs}). This is further discussed below.

\begin{figure}
\centering
\includegraphics[width=.483\textwidth]{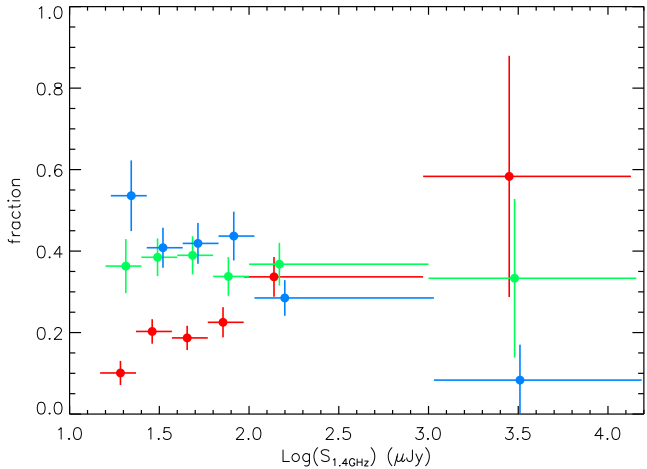}
\caption{The relative contributions of different SED--classified
  samples as a function of radio flux density. The points show the
  median 20cm flux density in the flux bin (indicated by the horizontal
  error bar) against the fraction over the whole population in that
  flux bin.  Color coding is the same as in figure \ref{fig:U-BSEDs}.
  The flux density bins for the different SED--classified samples are
  the same, but are shown with a small offset in the figure for
  clarity.
\label{fig:zdist_subsamples}}
\end{figure}

The different (sky--coverage corrected) contributions of the red,
green, and blue classified sources are shown in figure
\ref{fig:zdist_subsamples}. This figure shows how the
$S_{1.4GHz}\geq100\mu$Jy sources, which have a relatively flat
redshift distribution in the redshift range we are probing (fig.
\ref{fig:zdist}), are roughly equally split between the three
sub--classes of red, green and blue galaxies.  When going to fainter
flux densities, where we saw that the redshift distribution becomes
more skewed toward higher redshifts, figure \ref{fig:zdist_subsamples}
shows that the sample is depleted of red galaxies while the
fraction of blue actively star--forming systems increases
(green--classified objects make up $\approx$30-40\% in all
sub--samples). In the faintest subsample, with $S_{1.4GHz}$ in the range
16--25$\mu$Jy, less than 25\% of the $0.3<z<1.3$ sample is at
$0.3<z<0.7$, compared to more than 40\% in the shallow
$S_{1.4GHz}\sim100\mu$Jy subsample; the faintest sample only hosts
$\sim$10\% red galaxies, while the fraction of blue galaxies
has increased to more than 50\%.  According to a
Kolmogorov--Smirnov test for these radio--faintest and brightest
samples, the U-B color distributions are different at a $>$99.7\%
significance level.

We also note that, by splitting host galaxies just based on their U-B
restframe color (e.g., red galaxies with U-B$>$1, blue galaxies with
U-B$<$1) we find that, in agreement with previous work
\citep[e.g.][]{mainieri2008}, the radio population is dominated by red
galaxies above flux densities of 100$\mu$Jy, while below 80$\mu$Jy
blue galaxies begin to dominate ($\approx 65$\%, considering a
flux--limited sample with $40<S_{1.4GHz} < 80 \mu$Jy). Splitting this
sample in three redshift bins with $0.3<z<0.6$, $0.6<z<0.9$ and
$0.9<z<1.3$, we find that this $\approx 65$\% fraction stays basically
constant as a function of redshift. However, we should think in terms
of luminosities instead of flux densities: while the flux density
range $40<S_{1.4GHz}<80 \mu$Jy at $z\sim 0.3$ corresponds to a
luminosity range of approximately 1--3$\times$10$^{22}$W/Hz, at $z\sim
1.3$ it corresponds to luminosities of order
3--6$\times$10$^{23}$W/Hz. These $\sim$10$^{23}$W/Hz luminosities in
the lowest redshift bin would correspond to flux densities well above
100$\mu$Jy, which as we said are usually dominated by red
galaxies. Therefore, the apparently mild evolution between redshift
$z\sim0.5$ and $z\sim1.1$ of the fraction of blue galaxies in
$S_{1.4GHz}<80 \mu$Jy samples actually happens against a change in
radio luminosities of an order of magnitude. This is due to evolution
in these faint populations as we will further discuss below (section
\ref{sec:LF}).

In the following we compare our SED--selected subsamples in terms of
different classification criteria from previous studies. 

\begin{figure}
\centering \includegraphics[width=.49\textwidth]{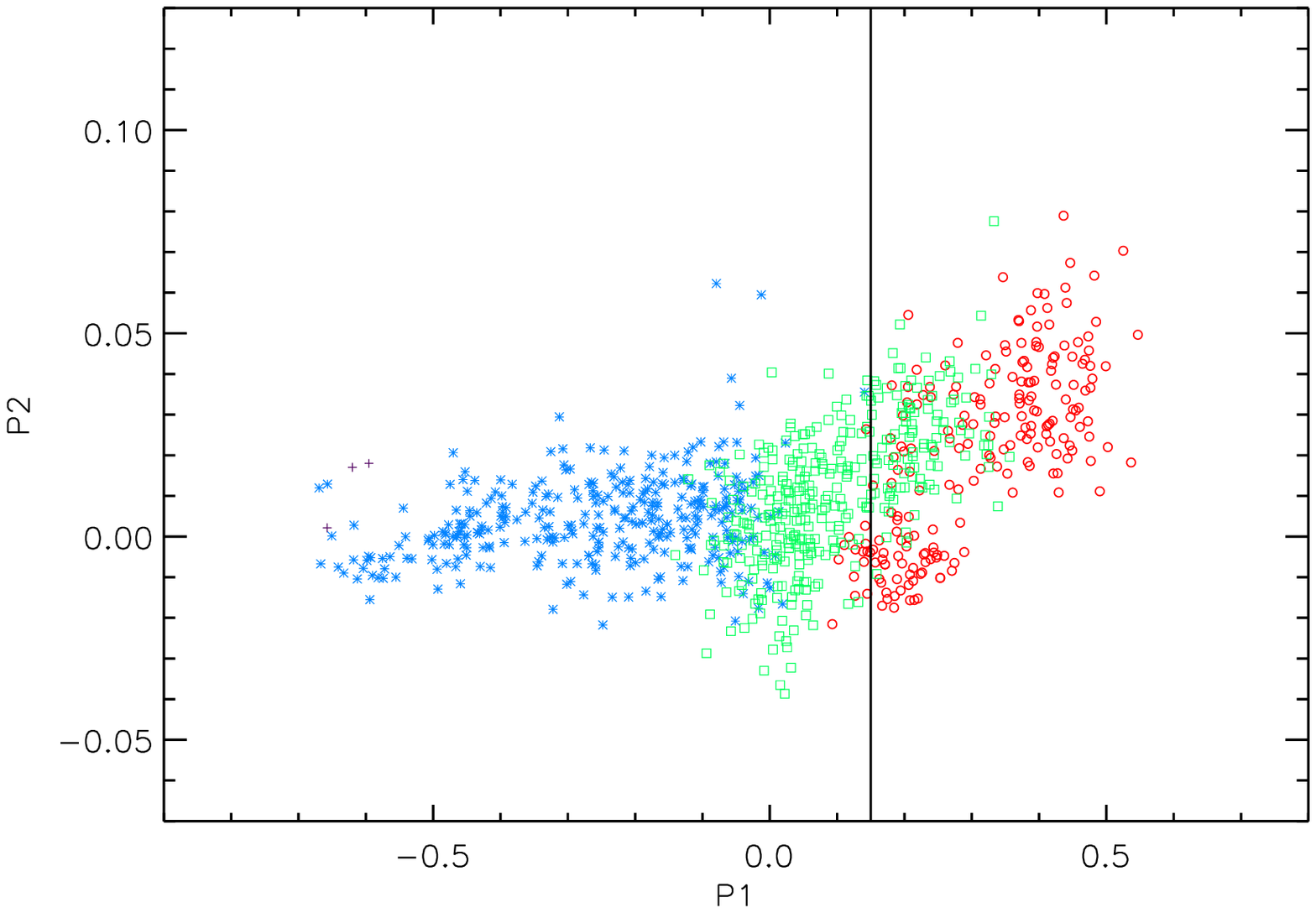}
\includegraphics[width=.49\textwidth]{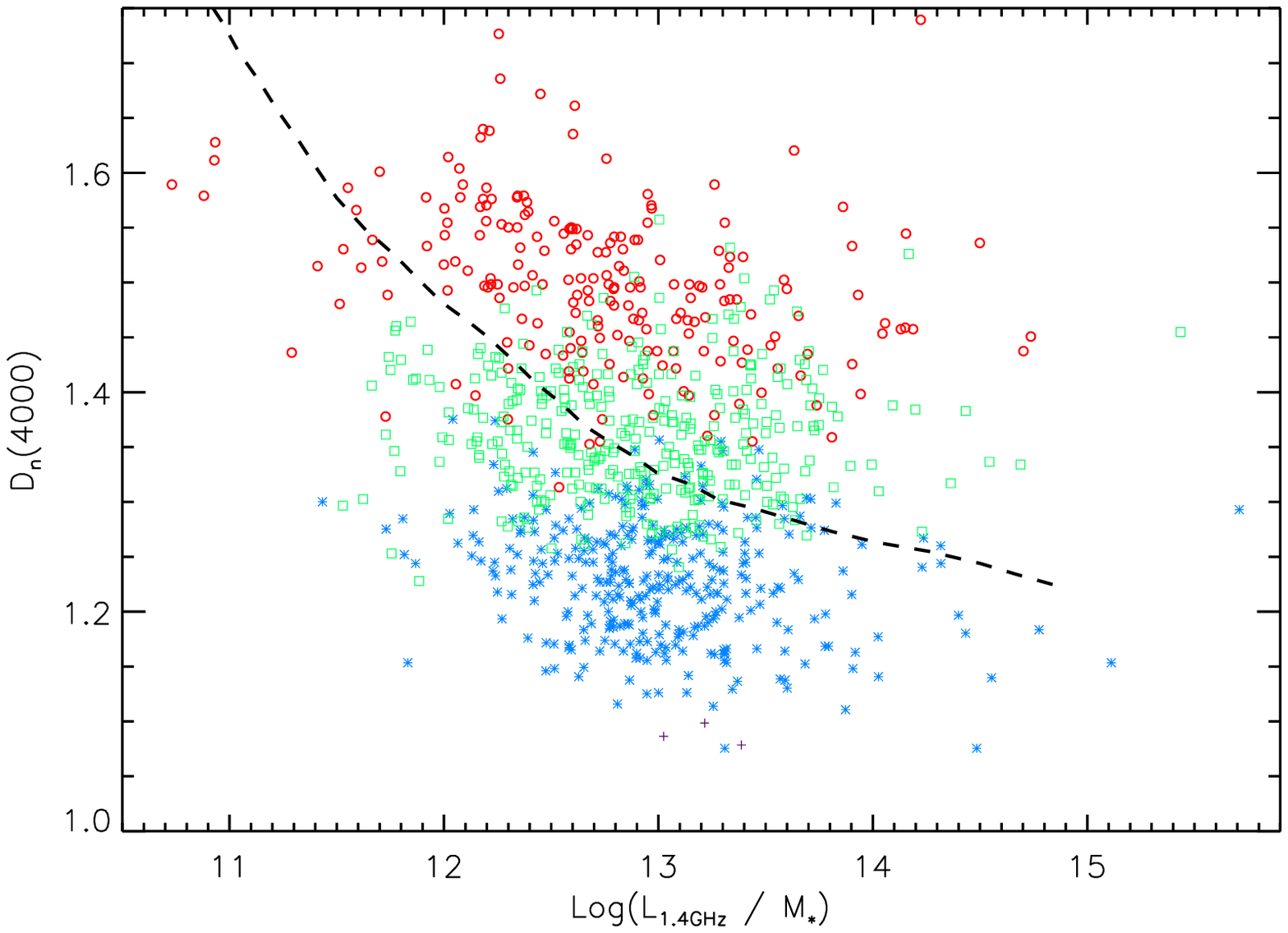}
\caption{{\bf Top panel:} The principal restframe color--color diagram
 (P1,P2) for objects with redshift $0.3 \leq z \leq 1.3$.  Symbols and
 color coding are based on the best--fit SED type as in figure
 \ref{fig:U-BSEDs}.  The solid black line shows the color--cut P1=0.15
 between AGN and star--forming dominated populations used in
 \citet{smolcic2008}.  {\bf Bottom panel:} The $D_{n}(4000)$ index
 plotted against the radio luminosity per unit stellar mass, for
 objects with redshift $0.3 \leq z \leq 1.3$. Symbols and color coding
 as in figure \ref{fig:U-BSEDs}. The dashed line shows the separation
 between AGN and star--forming galaxies as defined in
 \citet{best2005}. For the purpose of this plot stellar masses were
 scaled to a \citet{kroupa2001} IMF.
\label{fig:p1p2dn4000}}
\end{figure}

\subsubsection{NUV/optical vs radio properties}

For the sake of comparison, we show in figure \ref{fig:p1p2dn4000} how
our classification based on the observed broad--band SED compares to
two different classifications of radio samples based on specific
rest--frame properties.  In the top panel of figure
\ref{fig:p1p2dn4000} we compare with the AGN/star--forming
classification used in \citet{smolcic2008}. Based on a SDSS/NVSS/IRAS
sample of local radio sources, \citet{smolcic2006,smolcic2008} devised
a classification method to separate populations of radio sources whose
1.4GHz emission is dominated by AGN, by star formation, or is likely
to be contributed by both processes. This is based on the principal
component restframe colors P1 and P2, which are linear combinations of
restframe colors in the modified Str\"{o}mgren system in the
wavelength range 3500 -- 5800 $\AA$ \citep{odell2002,smolcic2006}. In
\citet{smolcic2008}, a color cut at P1=0.15 was adopted to separate
their radio sample in the COSMOS field into two populations of AGN and
star--formation dominated systems.  As the figure shows, indeed also
our objects with a P1$>0.15$ are mainly classified as quiescent (thus
likely with an AGN produced radio emission), or at most as
intermediate color sources (thus with a possibly relevant contribution
by AGN). On the other hand, our class of intermediate objects extends
down to slightly below P1=0, where increasingly star forming
populations become dominant extending down to P1$\approx -0.7$. Thus,
to first approximation, most of our green objects would be classified
as star--forming in this scheme\footnote{We note that
\citet{smolcic2008} applied a correction to the synthetic P1 color
used for the COSMOS sample, through comparison of the P1 colors
estimated by the SED-fit model and by the spectrum for a sample of
SDSS sources. No such correction was applied here. If applying to our
P1 colors the same corrections used in \citet{smolcic2008}, the P1
color of green objects would be on average lower by about 0.06 mag,
thus moving further into the range of sources classified as
star--forming.}. Nonetheless, from e.g. \citet{smolcic2006}, we note
that a significant fraction of the sources with P1 around zero
(roughly $-0.1<$P1$<0.1$) populates the region in the BPT
\citep{BPT81} diagram where the AGN/star--forming classification is
considered uncertain.

In the bottom panel of figure \ref{fig:p1p2dn4000} we plot instead our
sample in the Log(L$_{1.4GHz}/M_{*}$) vs $D_{n}(4000)$
plane. \citet{best2005} divided a sample of SDSS/NVSS matched sources
at redshift $z<0.3$ into two broad classes of AGN and starbursting
galaxies. The selection was in fact based on the 4000$\AA$ break index
$D_{n}(4000)$ and radio luminosity per unit stellar mass
L$_{1.4GHz}/M_{*}$. Since we don't have spectra for all sources in our
sample, we use synthetic $D_{n}(4000)$ indices derived from best--fit
stellar population synthesis models, which may only be considered as a
very approximate estimate of the real $D_{n}(4000)$. We use the
stellar masses described above to calculate L$_{1.4GHz}/M_{*}$ for
each galaxy.  {\it If} the estimates we plot are representative enough
of the true $D_{n}(4000)$ and L$_{1.4GHz}/M_{*}$, we should conclude
that indeed basically all of our blue objects are below the
\citet{best2005} division line, and the vast majority ($>$90\%) of the
red galaxies are above the line\footnote{As far as these minor
differences are concerned, we note that the exact shape of the
division line was determined, also based on emission--line
diagnostics, for a local ($z \leq 0.1$) galaxy sample, while the
galaxies in our sample were observed 2 to 7 billion years earlier, as
it is clear from the $D_{n}(4000)$ range.}. The green sources fill the
gap between the two, just about the division line, falling in both the
``AGN'' and ``starburst'' regions.

\begin{figure}[ht!]
\centering
\includegraphics[width=.483\textwidth]{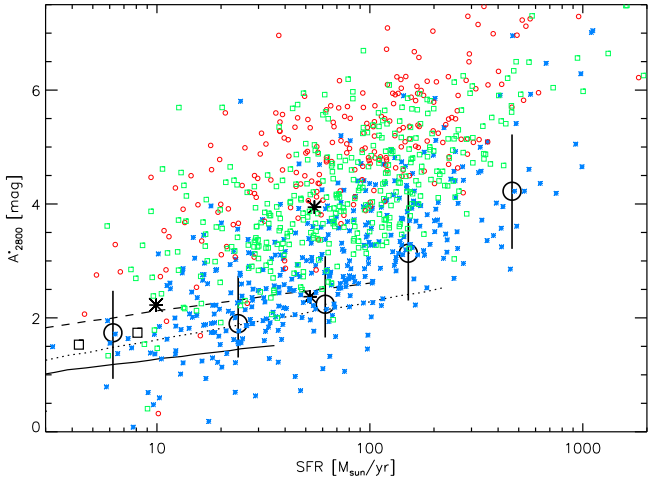}
\caption{The ``dust attenuation''
  A$^{*}_{2800}$=2.5Log(SFR$_{radio}$/SFR$_{2800\AA}$) against ``star
  formation rate'' as determined from the 1.4GHz luminosity (see text
  for the actual meaning of the two quantities plotted). Small symbols
  color coded as in figure \ref{fig:U-BSEDs} show the $0.3<z<1.3$
  radio sample (21 objects (out of which $>$95\% are red or green)
  with A$^{*}_{2800}>7.5$ and/or SFR$_{radio}>2000$ are not shown).
  Large empty circles with errorbars show the \citet{pannella2009}
  results (median and 16-84$^{th}$ percentile range) for star forming
  BzK galaxies at $z\approx 2$. Empty squares and stars show local
  galaxies from \citet{calzetti2000,calzetti2001} (star symbols are
  for galaxies classified as dust--rich in \citet{calzetti2001}; note
  NCG 1614 with SFR=55M$_{\odot}$/yr and very high dust extinction,
  which by the way hosts an obscured AGN (e.g.,
  \citet{guainazzi2007})). The solid, dotted and dashed lines show the
  dust attenuation as a function of SFR as derived from the local
  optical/UV--selected sample of \citet{hopkins2001}, the intermediate
  redshift NIR/MIR--selected sample of \citet{choi2006} and the
  low--redshift radio--selected sample of \citet{afonso2003},
  respectively (see text for details).
\label{fig:A1500}}
\end{figure}

Finally, we show in figure \ref{fig:A1500} how our classification of
this sample compares with other galaxy populations with regard to
radio/UV flux densities. Figure \ref{fig:A1500} shows for our sample
the star formation rate (SFR), as determined from the 1.4GHz
luminosity, against the dust attenuation estimated as
A$^{*}_{2800}\equiv$2.5Log(SFR$_{radio}$/SFR$_{UV}$). We note that the
meaning of these quantities for the whole sample is ill defined. In
fact, only for galaxies whose radio emission is due to star formation
the quantity plotted as ``SFR'' actually represents the SFR, and
A$^{*}_{2800}$ indeed is the dust attenuation. Instead, galaxies 
hosting a radio emitting AGN have their 1.4GHz luminosity at least
partially contributed by the AGN, and thus both ``SFR'' and
A$^{*}_{2800}$ lose their meaning for these objects. However, keeping
this in mind, this plot allows us to compare in a simple way our
sample with other populations of star forming galaxies at different
redshifts and selected with different criteria\footnote{ We note that
the plotted A$^{*}_{2800}$ and SFR values from other studies have been
derived from the original published quantities with some assumptions,
namely: 1) the {\it corrected} SFRs derived from 1.4GHz, H$_{\alpha}$,
UV or IR luminosities agree with each other; 2) the dust obscuration
estimated from the comparison of radio or IR vs UV fluxes, UV slope,
or Balmer decrements, also agree with each other once the appropriate
translations are made; 3) the \citet{calzetti2000} law; and 4) the
color excess for the stellar continuum is a factor 0.44 of the color
excess for the nebular gas emission lines.}.

As the figure shows, the range of attenuations derived for our blue
objects is in very good agreement with other studies of different
kinds of star forming galaxies at different redshifts
\citep{calzetti2000,calzetti2001,hopkins2001,afonso2003,choi2006,pannella2009}. While
part of the green objects would also overlap with these samples, it is
clear how the A$^{*}_{2800}$ for the green population generally lies
above the expectations, and definitely the red population has too high
values of A$^{*}_{2800}$. This might support the idea that the red
population is for the great majority made of AGN hosts, and also
suggest that at least part of the green population has a contribution
from AGN to the radio luminosity, even though part of these galaxies
can still be very dusty systems.

The significant occurrence of AGN hosts among the intermediate
population between red and blue galaxies has been noted in several
previous studies \citep[e.g.,][and references
therein]{choi2009,martin2007,nandra2007}, as well as their actual
nature of composite (meaning SF+AGN) systems
\citep{schawinski2007,wild2007,salim2007}.  We note that the
classification of such composite sources in the literature is quite
variable, and for instance galaxies with more than 10\% of their radio
luminosity contributed by star formation have been classified, in some
cases, as starbursts \citep{tasse2008}. This may indeed be the case,
and certainly also in our green sample different amounts of star
formation are presents, however a starforming--or--AGN classification
may be not appropriate for these sources, especially depending on the
kind of study they are used for.

\subsubsection{Radio--IR properties}

As discussed above, in spite of the insight that we can certainly gain
by combining radio fluxes and optical/NIR photometry, the results
presented so far may not provide conclusive evidence about the actual
nature of these sources.  Therefore, we need to introduce further
information which may help us identify which process powers the radio
emission. Obvious promising data already available on this field are
X--ray (Chandra) and infrared (Spitzer and Herschel) observations.
While we postpone a full analysis of these data to a future work, we
use here just the Spitzer/MIPS 24$\mu$m data to estimate the total
infrared (IR) luminosity of our sources and thus examine the behavior
of our SED--selected sub--samples with respect to the FIR--radio
correlation \citep[e.g.,][]{condon1992,yun2001}. The DSF was observed
at 24$\mu$m with MIPS onboard Spitzer as part of the GO--3 program
\#30391 (PI: F. Owen), for a total of 60.6 hours over an area of about
half square degree, and a median integration time per pixel of about
2500s. The $5\sigma$ flux density limit is estimated to be about
40$\mu$Jy. The data reduction and catalog production are described in
full detail in a forthcoming companion paper (Owen et al., in
preparation). More than 80\% of the faint radio sources from the 90\%
complete sample in the redshift range $0.3<z<1.3$ are detected at
24$\mu$m. Matched sources with flux {\it possibly} contaminated by
neighbors within the 24$\mu$m PRF are estimated to be about 10\% of
all matched detections, based on the cross--correlation with the IRAC
3.6$\mu$m catalog.  While upper limits make up for $<20\%$ of the
whole $0.3<z<1.3$ sample, they are more relevant for red--classified
objects (almost 40\% of upper limits) than for green and blue sources.

In the redshift range $0.3<z<1.3$, the observed 24$\mu$m light probes
  the restframe $\sim10-20\mu$m. We use the templates of
  \citet{charyelbaz2001} to estimate from the 24$\mu$m flux density
  the total (8-1000$\mu$m) IR rest--frame luminosity (or an upper
  limit for 24$\mu$m--undetected sources). This is done taking the
  template whose predicted luminosity at the {\it observed 24$\mu$m} is closest
  to the actual observed 24$\mu$m luminosity. Taking the mean or median
  luminosity across the whole template library typically changes this
  number within about 0.15dex, but since this is obviously
  library--dependent we don't see a real point of adopting the mean or
  median instead of the formal best--fitting template. Furthermore,
  even just based on the templates available in this library, such
  estimate of the total IR luminosity based on the single observed
  24$\mu$m point may be affected by systematics of up to a factor 2.

\begin{figure}[hb!]
\centering
\includegraphics[width=.483\textwidth]{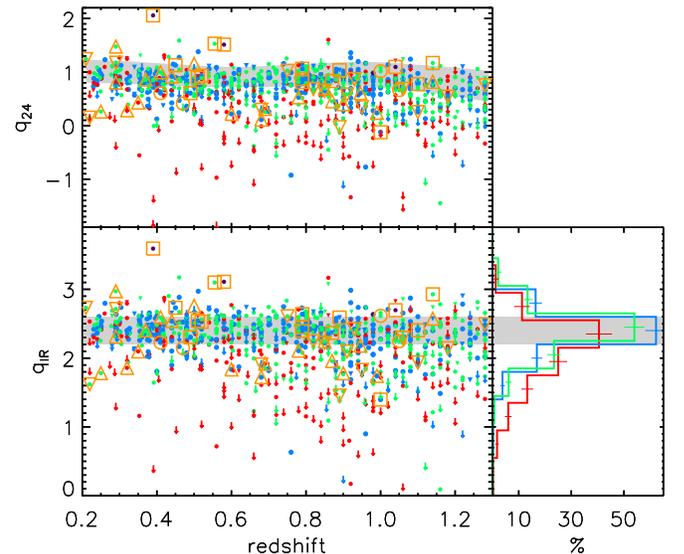}
\caption{{\bf Bottom panels:} in the left panel the flux ratio
$q_{IR}$ is plotted against redshift for the different SED--selected
sub--samples (color coding as in figure \ref{fig:U-BSEDs}). Filled
circles show unambiguously matched $24\mu$ sources, filled triangles
show matched $24\mu$ sources whose flux might be contaminated by
neighbors, and arrows show upper limits (see text for details). Large
orange symbols show matched X--ray sources from the
\citet{trouille2008} sample (according to the \citet{trouille2008}
spectral classification, circles, downward triangles, triangles and
squares show absorbers, star formers, high--excitation sources and
broad--line AGNs, respectively). The gray--shaded area shows the
$1\sigma$ range about the (redshift independent) $q_{IR}=2.4$ reported
in \citet{ivison2009}. The right-hand panel shows the sky--coverage
corrected distributions of $q_{IR}$ for the sub--samples plotted in
the left panel, including upper limits (see text for details). The
histograms for the red, green, and blue samples are evaluated in the
same $q_{IR}$ bins, but are shown slightly offset for clarity. {\bf
Top panel:} same as bottom left panel, but for the {\it observed} (non
k--corrected) flux density ratio $q_{24}$.  The gray shaded area shows
the envelope of \citet{charyelbaz2001} and \citet{rieke2009} templates
with $L_{IR}$=10$^{11}L_{\odot}$ and 10$^{12}L_{\odot}$ (plus 
10$^{10}L_{\odot}$ for $z\lesssim 0.5$).
\label{fig:qir}}
\end{figure}

We use this total IR luminosity together with the 1.4GHz luminosity to
estimate the logarithmic ratio of bolometric IR and monochromatic
radio luminosities $q_{IR}$=Log($L_{IR}/(3.75\cdot10^{12})$) -
Log($L_{1.4GHz}$) \citep{helou1985}. This is plotted as a function of redshift
for the different sub--samples of red, green, and blue sources, in the
bottm-left panel of figure \ref{fig:qir}. In this figure, filled circles
show sources unambiguously matched with a 24$\mu$m source, and filled
triangles show sources matched within 2'' with a 24$\mu$m source but
whose 24$\mu$m flux {\it might} be contaminated neighbors. Finally,
upper limits are shown by down--pointing arrows.

The bottom-right panel of this figure shows the $q_{IR}$ distribution
(corrected for the sky--coverage) for the three different red, green,
and blue sub--samples.  It is important to note that the plotted
histograms (as well as the related statistics given below) include
both 24$\mu$m detections and upper limits. This does not affect our
results (or actually affects our conclusions in a conservative way),
since we are interested in the difference between the $q_{IR}$
distribution of the three sub--samples, with $q_{IR}$ of intermediate
and quiescent sources expected to be lower than $q_{IR}$ of star
forming sources, if the radio power of objects classified as
intermediate and quiescent is at least partially provided by an AGN.
This means that, including upper limits (and possibly contaminated
24$\mu$m detections), we are -- if anything -- {\it reducing} the
actual difference between the $q_{IR}$ distributions of the three
sub--samples.

In both panels, the gray--shaded area shows the $1\sigma$ range about
the (redshift independent) $q_{IR}=2.4$ reported in
\citet{ivison2009}.  As figure \ref{fig:qir} shows, blue star--forming
sources mostly lie on the expected FIR--radio correlation. Also up to
50\% of the green sources lie within $1\sigma$ of the expected
FIR--radio correlation, which would point toward these being powered
by star formation as well. However, even though with a broader
distribution, red sources also lie close to the FIR--radio
correlation, with up to 40\% of this sub--sample lying within
1$\sigma$ of the correlation. Sources populating the gray--shaded area
in figure \ref{fig:qir} (1$\sigma$ about the expected correlation) are
for more than 50\% classified as star--forming, but show a sizable
fraction of $\sim30\%$ and $\sim10\%$ of green and red sources,
respectively.  The $q_{IR}$ distribution of red sources appears to be
very different from those of blue and green sources. Even though a
Kolmogorov-Smirnov test cannot be meaningfully applied to the
sky--coverage corrected distributions, applying it to flux--limited
sub--samples selected in portions of the radio image with depth
uniformly better than a given threshold, suggests than the $q_{IR}$
distribution of red sources is different at a significance of more
than $99.9\%$. A $\chi^{2}$ test on the binned distributions
(including errors) plotted in figure \ref{fig:qir} (as well as on
similar distributions binned with half bin size), also suggests that
the distributions of $q_{IR}$ of red vs. blue or green sources are
different at a $>99.5\%$ level.  On the other hand, the $q_{IR}$
distributions of blue and green sources look much more similar. The
$\chi^{2}$ test on the binned distributions suggest that they are
different at a $\gtrsim 98\%$ level, and the Kolmogorov-Smirnov test
on a limited part of the sample (as above) gives a $P_{KS}\sim 0.2$,
thus based on the present data these distributions are, at most,
marginally different.

Based on these results one might thus conclude that a large fraction
of our sources with intermediate colors, as well as a sizable fraction
of the red ones, are actually reddened starbursts. On the other hand,
we also note that among the sources which are X--ray detected (and
mostly AGN classified in the \citet{trouille2008} sample), many lie on
the FIR--radio correlation, including some classified as absorbers
(based on OII or H$_{\alpha}$+NII equivalent widths). Indeed, other
studies have found that sources classified as low--radio power AGN may
often lie on or close to the same FIR--radio correlation expected for
star forming sources \citep[e.g. see discussions and references
in][]{smolcic2008,sargent2010}.  However, it should also be noted
that an AGN selected based on its (spectral or photometric)
optical/NIR properties (or X--ray), hosted in a ``composite'' system
with ongoing star formation, may be in a stage where radio emission is
negligible and thus does not significantly affect the IR/radio
properties which remain determined by the star formation process. In
this case, while information at other wavelengths suggest the presence
of an AGN, it might not (significantly) contribute to the radio power
of the source. Furthermore, obscured star formation confined in
limited areas of a galaxy might also determine its IR/radio properties
while possibly going undetected in broad--band photometry at optical
wavelengths.  On the other hand, we remind again the reader that the
$q_{IR}$ in figure \ref{fig:qir} relies uniquely on the observed
24$\mu$m flux density, and thus might be biased producing a spurious
result. For comparison, we also show in the top panel of figure
\ref{fig:qir} the observed (non k--corrected) flux density ratio
$q_{24}=Log(S_{24\mu m}/S_{1.4GHz})$. As a reference, the gray shaded
area shows the envelope of \citet{charyelbaz2001} and
\citet{rieke2009} templates with $L_{IR}$=10$^{11}L_{\odot}$ and
10$^{12}L_{\odot}$ (plus also 10$^{10}L_{\odot}$ for $z\lesssim
0.5$). This $q_{24}$--based figure essentially reflects the results of
the bottom panel. We will continue the investigation of IR/radio
properties of this sample with a proper SED analysis of the full data
in a future work.

\begin{figure}
\centering
\includegraphics[width=.49\textwidth]{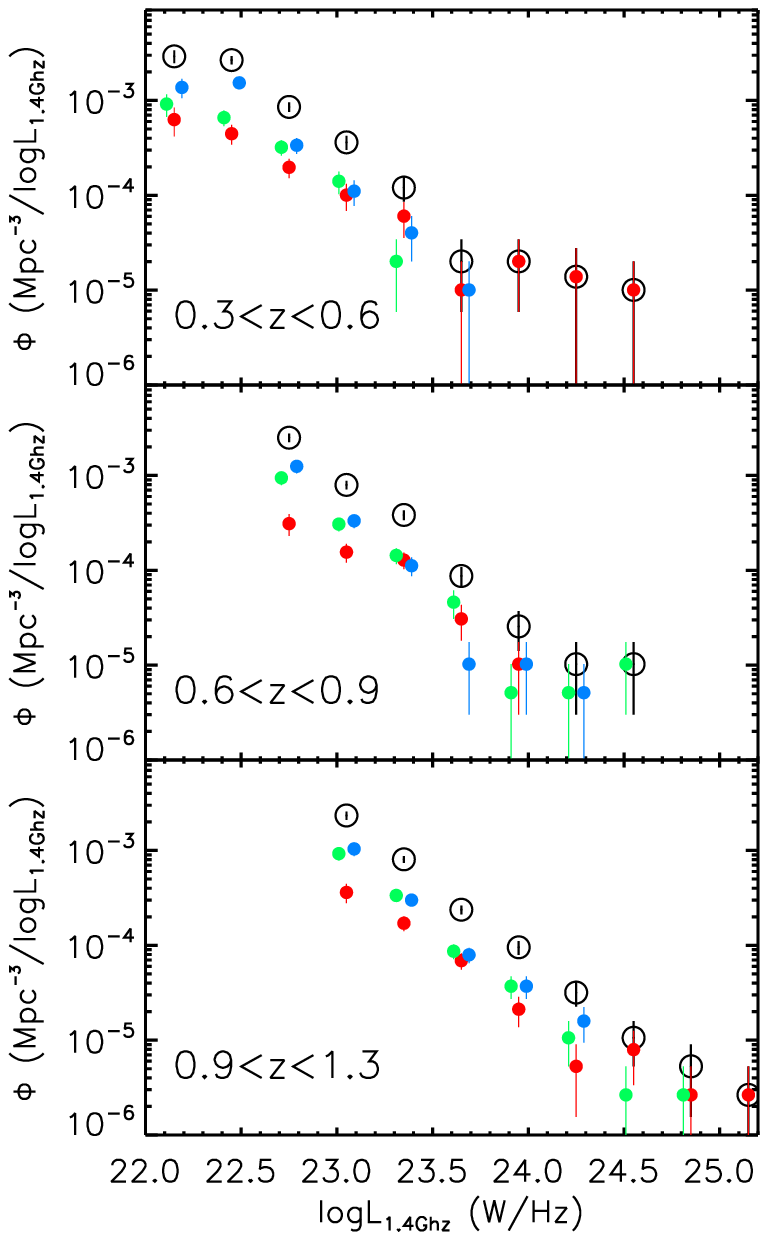}
\caption{The 1.4GHz luminosity function of different galaxy
  classes.  Large empty symbols refer to the whole populations, while
  filled symbols and their coding refer to the same classes as in
  figure \ref{fig:U-BSEDs}. Number densities are shown in three
  redshift ranges, as indicated. The luminosity functions for the
  different samples are evaluated in the same $L_{1.4GHz}$ bins, but are
  shown slightly offset for clarity. Errorbars show Poissonian
  errors.\\
\label{fig:LF}}
\end{figure}

\begin{figure}
\centering
\includegraphics[width=.483\textwidth]{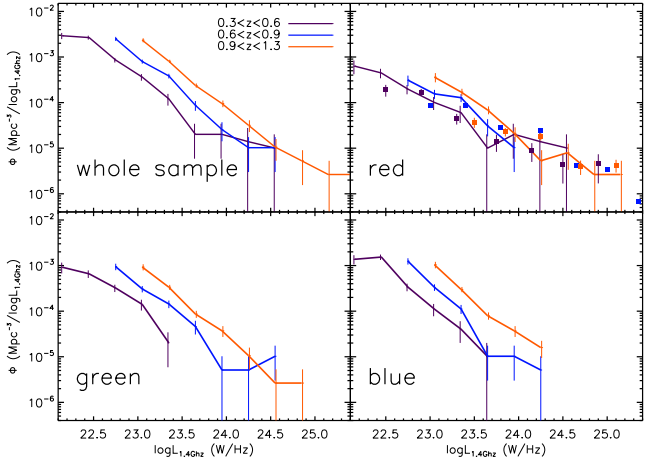}
\caption{The redshift evolution of the 1.4GHz luminosity function
  of galaxies in different SED classes, as indicated in each panel.
  Luminosity functions are the same as those plotted in figure
  \ref{fig:LF}. Different colors refer to different redshift ranges,
  as also indicated. Errorbars show Poissonian errors. Small squares
  (with same color coding as solid lines) show, as a reference, the LF
  of radio--selected AGN as derived by \citet{smolcic2009b} in similar
  redshift bins (see text for details).
\label{fig:LFzevol}}
\end{figure}

\subsection{Luminosity functions and evolution of faint radio populations}
\label{sec:LF}

Finally, we investigate the contributions of the different galaxy
classes to the sub--mJy population by plotting in figure \ref{fig:LF}
the 1.4GHz luminosity functions (total and split by SED class) in
three redshift ranges. Figure \ref{fig:LF} shows number densities
based on the $0.3<z<1.3$, $90\%$ complete sample, estimated with the
1/V$_{max}$ method \citep{schmidt1968,avni1980}. Within the redshift
range $0.3<z<1.3$, the maximum volume over which an object can enter
our sample essentially depends on its radio luminosity and on the (non
uniform) depth of the 1.4GHz image. Therefore, for each source of a
given radio luminosity $L_{1.4GHz}$, the maximum volume accessible to
the source was calculated, based on the maximum redshift out to which
the source would have been detected as a function of the varying image
depth of the 1.4GHz image. We remind the reader that, similarly
to what discussed above concerning the sky coverage correction, the
1/V$_{max}$ correction is not negligible for low--luminosity sources
due to the limited survey area probing the faintest fluxes. On the
other hand, we also note here that luminosity functions obtained,
without the use of the 1/V$_{max}$ correction, by using smaller volume
limited samples defined in portions of the radio image and luminosity
ranges, are generally in very good agreement with those determined
with the 1/V$_{max}$ method, and lead to the same conclusions.

Only sources in the $>90\%$ complete sample were used, and no further
correction was adopted for the small residual incompleteness of the
sample. However, we remind the reader again that the $>90\%$
completeness level of this sample was estimated based on some
assumptions (as discussed in section 5.3).

Figure \ref{fig:LF} shows the relative contributions of the different
SED--classified sub--samples as a function of redshift and luminosity.
It also suggests an evolution of the luminosity functions of all
classes over the probed redshift range, which is shown more clearly in
figure \ref{fig:LFzevol}. This is confirmed by a $\chi^{2}$ fit to the
binned data in the three redshift bins with a parametric LF of the
\citet{saunders1990} form $\Phi(L)=\Phi^{*}
(L/L_{*})^{1-\alpha}${\small
exp}$(-0.5\sigma^{-2}[$log$(1+L/L_{*})]^{2})$. We note that, in order
to keep consistency within our SED classification, the fit was
performed only in the three redshift bins shown in figures
\ref{fig:LF} and \ref{fig:LFzevol}, and over the luminosity range
properly probed by our data (as a reference,
Log($L_{1.4GHz}$)$<24.1,24.4,24.6$ in the $z\sim0.5,0.8,1.1$ redshift
bins, respectively), without including measurements from other surveys
sampling brighter luminosities, nor a z=0 reference LF. Nonetheless,
we show in figure \ref{fig:LFzevol} the LF of radio selected AGNs as
derived by \citet{smolcic2009b} in the redshift bins $0.35<z<0.6$,
$0.6<z<0.9$, $0.9<z<1.3$. We note that the \citet{smolcic2009b} LFs are
shown just as a reference, as the AGN sample of Smolcic et al. does
not perfectly match our red sample, as shown in figure
\ref{fig:p1p2dn4000}, top panel\footnote{Because of the even greater
difference between our blue and green samples and the Smolcic et
al. star-forming sample we do not attempt a comparison of the LFs for
star--forming galaxies.}. Given this difference in the sample
selection (and a small difference in the first redshift bin), the
small area of our field which does not properly probe the brighter
luminosities better sampled by the large COSMOS survey, and in general
the uncertainties involved in the LF determination, our red--sample
LFs can be considered in reasonably good agreement with the AGN LFs of
\citet{smolcic2009b}.

For each SED class (red, green and blue), as well as for the total
population, we performed a simultaneous fit to the binned LFs in all
three redshift bins, allowing for redshift evolution in the form $L
\propto$ (1+z)$^{\alpha_{L}}$. This assumption of pure luminosity
evolution (PLE) is only used as a working tool in order to quantify
the observed evolution, and for comparison with other studies. In
fact, we are well aware of the fact that, beside the well known
degeneracy between luminosity and density evolution, there is indeed
little reason to believe that either pure luminosity or pure density
evolution (PDE) may be an adequate description over a range of
different luminosities and redshifts \citep[e.g.][]{dunlop1990,willott2001,ueda2003}. The very fact that
radio populations are made of different types of objects, and that
these different sub--populations might evolve in different ways, imply
that simple PLE or PDE cannot be in general considered as an adequate
description. On the other hand, due to the limited luminosity range
probed, and to the small number statistics, our data alone cannot be
sufficient to effectively constrain the general 20cm luminosity
function and its redshift evolution, so we will only try to quantify
the amount of evolution observed, at the luminosities and redshifts
properly probed by our data, by {\it assuming} the simple PLE model of
redshift evolution.

The best--fit $\alpha_{L}$ obtained for the whole population is $3.5
\pm 0.2$, while for the red, green, and blue sub--samples we obtain
$\alpha_{L}$=2.7$\pm0.3$, 3.7$^{+0.3}_{-0.4}$, and
$3.2^{+0.4}_{-0.2}$, respectively. A non--parametric evaluation of
$\alpha_{L}$, obtained by directly comparing the LFs in the three
redshift bins (and assuming $L \propto$ (1+z)$^{\alpha_{L}}$ as
above), instead of fitting the parametric \citet{saunders1990} form,
yields results perfectly consistent with those listed above for the
parametric fitting ($3.1\pm0.2$, $2.5\pm0.3$, $3.6\pm0.2$,
$2.9\pm0.3$, for the total, red, green, and blue samples,
respectively).

The formal best--fit $\alpha_{L}$ is thus close to $\approx 3$ for the
whole sample as well as for all SED sub--samples.  This suggests that
our observations are consistent with luminosities decreasing by a
factor $\approx 10$ from redshift just above 1 to the local Universe.
The fact that the evolution factors for the three SED classes are very
close to each other might suggest also in this work a link between the
evolution of star formation and AGN activity \citep[e.g.,][and references therein]{silverman2008,heckman2009}, provided
that our SED--selected sub--samples actually probe such different
populations.

A PLE rate $\alpha_{L} \approx 3$ is similar to PLE rates estimated in
other studies for star--forming galaxies (e.g., 2.5
\citep{seymour2004}, 3 \citep{cowie2004}, 2.7 with a negligible
$\alpha_{D}$=0.15$\pm0.6$ \citep{hopkins2004}, 2.7 \citep{huynh2005},
2.3 \citep{moss2007}, 2.1-2.5 \citep{smolcic2009}) or X--ray AGNs
(e.g., 3.2 \citep{barger2005}, 2.7 \citep{dellaceca2008}).  As far as
AGNs are concerned, we should note though that low--luminosity radio
AGNs have been found to show slower evolution compared to higher radio
power AGNs (e.g., \citet{willott2001}) and, at luminosities similar to
those probed here, somewhat lower PLE rates as compared to our red
sample have been measured in some previous work (e.g. $\alpha_{L}$=2
\citep{sadler2007} or $\alpha_{L}$=0.8 \citep{smolcic2009b} using the
\citet{sadler2002} AGN LF).

\section{Summary}

We have carried out a multi--wavelength analysis of faint radio
sources in the Deep Swire VLA Field. The depth of this survey allows
us to probe populations of radio sources uncommon or absent in other
deep radio surveys, with almost a thousand sources fainter than
50$\mu$Jy.

Based on optical/NIR/IRAC photometry, we built the SEDs of the
identified counterparts and compared them both with a galaxy SED
template library and with stellar population synthesis models,
determining their photometric redshifts, stellar masses, and broad
stellar population properties. The derived redshift distribution of
radio sources appears to be different at different flux density
levels, with the distribution of the faintest sources showing a more
pronounced peak, at about redshift one.

We have focused on a $90\%$ complete sample of counterparts of
sub--mJy radio sources with redshift $0.3<z<1.3$, dividing the sample
in broad classes of quiescent, intermediate and star--forming systems
based on their optical/NIR colors. The population mix as described by
these sub--samples shows a clear dependence on radio flux density,
with an increase of star--forming populations at lower fluxes, in
agreement with previous studies.  At all redshifts up to $z\sim 1.5$,
the contribution of star--forming galaxies becomes increasingly
important at lower radio luminosities.

The rest--frame U-B vs B color--magnitude diagram of this
radio--selected sample shows the presence of a significant
green--valley population, beside two populations of red and blue
galaxies. One might assume that the radio emission from red galaxies
is in most part due to an AGN, because of their apparently very low
star formation (beside a possible contribution of extremely dusty
galaxies), and on the other hand that there is a predominant
contribution to the radio emission from star formation in blue star
forming galaxies.

The properties of the intermediate, green valley sample with respect
to the ``AGN'' vs ``star--forming'' classification are less
clear. Their stellar population properties, and the comparison of
radio and UV luminosities, suggest that this class of objects might be
a mixed population of AGN and star--forming galaxies, possibly
including composite systems where both nuclear activity and star
formation are present.  On the other hand, the comparison of
24$\mu$m--based IR and 1.4GHz radio luminosities would suggest that
many of our faint radio sources, and not limited to the blue,
obviously star--forming galaxies, may lie close or on top of the
radio--IR correlation expected for star forming objects. In
particular, not only the radio--to--IR flux ratios of sources
classified as intermediate are distributed very similarly to those of
star--forming galaxies, but also up to 40\% of the ``quiescent''
classified galaxies have a measured $q_{IR}$ in the range 2.2--2.6.
This result may be affected by the fact that we are sampling the IR
SED only with one (24$\mu$m) photometric point.  More stringent
conclusions will be possible thanks to the analysis of the full IR SEDs, which will be the
subject of a forthcoming work.

In agreement with previous studies, all populations studied in this
work show evolution with redshift. In a simple PLE scenario, 1.4GHz
luminosities decrease by about a factor 10 from redshift just beyond
one to the local Universe. 

A comparison with the whole sample of
radio--undetected objects in the field, as well as a stacking analysis
to study average radio properties of complete mass/SED--selected
galaxy samples, are postponed to a future work, and will likely lead
to a better understanding of the nature of the populations studied
here, as well as more in general of the interplay of AGN and star
formation activity in the evolution of galaxies.

\acknowledgments We are grateful to Niv Drory for sharing the
SED--fitting code used to estimate galaxy stellar masses. We thank the
referee, Paolo Padovani, for his careful reading of this manuscript,
and for his detailed comments and suggestions which helped us improve
the clarity and presentation of this work.  Based on observations of
the SWIRE Lockman Hole field taken on the NRAO VLA, Spitzer Space
Telescope, KPNO Mayall telescope, UKIRT and CFHT.  The National Radio
Astronomy Observatory is a facility of the National Science Foundation
operated under cooperative agreement by Associated Universities,
Inc.. The Spitzer Space Telescope is operated by the Jet Propulsion
Laboratory, California Institute of Technology under a contract with
NASA. The United Kingdom Infrared Telescope is operated by the Joint
Astronomy Centre on behalf of the Science and Technology Facilities
Council of the U.K.. The Canada-France-Hawaii Telescope (CFHT) is
operated by the National Research Council (NRC) of Canada, the
Institut National des Science de l'Univers of the Centre National de
la Recherche Scientifique (CNRS) of France, and the University of
Hawaii.  This research used the facilities of the Canadian Astronomy
Data Centre operated by the National Research Council of Canada with
the support of the Canadian Space Agency.  FNO and GEM were visiting
astronomers at the Kitt Peak National Observatory, National Optical
Astronomy Observatory, operated by AURA, Inc., under cooperative
agreement with the National Science Foundation. This work was
supported by NASA through Jet Propulsion Laboratory contract
No.1289215. VS and MP acknowledge support from the Max-Planck Society
and the Alexander von Humboldt Foundation. WHW acknowledges support
from NRAO through a Jansky Fellowship.

\bibliography{stra0720_rev2_emul}

\begin{table}[h!]
\caption{Values of the magnitudes mag$_{cut}$ and mag$_{10}$
  derived through simulations of photometric measurements in apertures
  of diameter 1.5\arcsec~ in dual--image mode of artificial
  point--like sources added to the images. The images marked with a
  * were smoothed to a FWHM=1.3\arcsec.  See text for details.
\label{tab:magcutcompl}}
\centering
\vspace{0.2cm}
   \begin{tabular}{@{}c c c @{}}
\hline
\hline
passband & mag$_{cut}$ &mag$_{10}$	\vspace{0.06cm}\\
 & (AB mag) & (AB mag)	\vspace{0.06cm}\\
\hline
NUV      &25.5  &24.0	\vspace{0.06cm}\\
U	 &28.0	&25.5	\vspace{0.06cm}\\
g$^{*}$	 &27.8	&25.3	\vspace{0.06cm}\\
r$^{*}$	 &27.3	&24.9	\vspace{0.06cm}\\
i$^{*}$	 &26.5	&24.3	\vspace{0.06cm}\\
z$^{*}$	 &26.2	&23.6	\vspace{0.06cm}\\
J$^{*}$	 &25.2	&23.0	\vspace{0.06cm}\\
H$^{*}$	 &25.0	&22.7	\vspace{0.06cm}\\
K$^{*}$	 &24.2	&22.4	\vspace{0.06cm}\\
3.6$\mu$m &24.4	&22.4	\vspace{0.06cm}\\
4.5$\mu$m &24.2 &22.2	\vspace{0.06cm}\\
\hline
\end{tabular}
\end{table}

\begin{table}
\caption{Some of the main derived properties used in this work. This
table is available in its entirety on the electronic edition, a
portion is shown here for guidance. Only objects with available
spectroscopic redshift or reliable photo--z are listed.  See text for
details about how these quantities were derived. Notes: ($^{a}$) ID, RA and Dec as in Paper I.
($^{b}$) Spectroscopic redshift where available, otherwise
photometric redshift. Redshifts without an error listed are spectroscopic.
($^{c}$) Quality flag for photometric
redshifts, as defined in section \ref{photozdet}.
($^{d}$) Synthetic estimates of absolute magnitude in the R
(Johnson) band, restframe U-B color, and stellar mass (Salpeter IMF),
assuming the redshift listed in column 4. Both R band magnitude and
stellar mass estimates are based on SED fitting in a 1.5\arcsec aperture
and corrected for each object based on the difference between total
and aperture magnitudes as measured in the detection image, as
explained in section \ref{mmll}.
\label{tab:bigtable}}
\centering
\vspace{0.2cm}
   \begin{tabular}{@{}c c c c c c c c @{}}
\hline
\hline
ID$^{a}$ & RA$^{a}$ & Dec$^{a}$ & redshift$^{b}$ & QF$_{\rm photo-z}^{c}$ & M$_{\rm R}^{d}$ & U-B$^{d}$ &
Log(M$_{*}$)$^{d}$\\ 
& (J2000) & (J2000) & & & (mag) & (mag) & (M$_{\odot})$ \\
\hline

     20032  &  10:43:41.58  &   59:09:20.4 & 1.66 $\pm$  0.50  &   AA &  -20.9  &  0.55 &  10.35 $\pm$   0.25	\vspace{0.06cm}\\
        31  &  10:43:43.59  &   59:00:10.2 & 0.92 $\pm$  0.05  &   AA &  -21.9  &  0.84 &  10.26 $\pm$   0.20	\vspace{0.06cm}\\
        35  &  10:43:43.93  &   58:56:51.6 & 0.64 $\pm$  0.07  &   AA &  -21.3  &  0.79 &  10.09 $\pm$   0.26	\vspace{0.06cm}\\
     20038  &  10:43:44.33  &   59:05:51.1 & 0.96 $\pm$  0.10  &   AA &  -21.1  &  1.01 &   9.96 $\pm$   0.32	\vspace{0.06cm}\\
        36  &  10:43:44.76  &   59:15:02.9 & 1.62 $\pm$  0.09  &   AA &  -24.2  &  1.09 &  11.00 $\pm$   0.22	\vspace{0.06cm}\\
        37  &  10:43:44.84  &   58:58:18.7 & 0.52 $\pm$  0.11  &   AA &  -22.8  &  1.17 &  11.24 $\pm$   0.13	\vspace{0.06cm}\\
     10056  &  10:43:45.22  &   58:57:40.2 & 2.08 $\pm$  0.24  &   AA &  -22.6  &  0.39 &  10.26 $\pm$   0.21	\vspace{0.06cm}\\
     20045  &  10:43:46.13  &   59:01:37.9 & 1.04 $\pm$  0.07  &   AA &  -21.6  &  0.93 &  10.24 $\pm$   0.39	\vspace{0.06cm}\\
        44  &  10:43:46.21  &   59:01:19.1 & 1.34 $\pm$  0.08  &   AA &  -23.9  &  1.01 &  11.02 $\pm$   0.14	\vspace{0.06cm}\\
        46  &  10:43:46.91  &   59:00:21.3 & 0.98 $\pm$  0.05  &   AA &  -21.9  &  0.98 &  10.35 $\pm$   0.36	\vspace{0.06cm}\\
     10061  &  10:43:47.91  &   59:06:21.2 & 0.14    &    - &  -19.3  &  0.57 &   9.60 $\pm$   0.26	\vspace{0.06cm}\\
        47  &  10:43:48.05  &   59:02:23.2 & 2.26 $\pm$  0.45  &    B &  -23.8  &  0.75 &  10.90 $\pm$   0.22	\vspace{0.06cm}\\
        49  &  10:43:49.22  &   58:55:38.7 & 1.58 $\pm$  0.33  &   AA &  -21.0  &  0.44 &   9.59 $\pm$   0.28	\vspace{0.06cm}\\
     20053  &  10:43:49.26  &   58:51:34.7 & 0.74 $\pm$  0.10  &   AA &  -22.0  &  1.10 &  11.03 $\pm$   0.18	\vspace{0.06cm}\\
        50  &  10:43:49.46  &   58:56:31.5 & 1.06 $\pm$  0.10  &   AA &  -22.5  &  0.97 &  10.69 $\pm$   0.18	\vspace{0.06cm}\\
\hline
\end{tabular}

\end{table}

\end{document}